\documentclass[%
 reprint,
 aps,
]{revtex4-1}
\usepackage[caption=false]{subfig}
\usepackage{ amsmath,amssymb}
\usepackage{epstopdf}
\usepackage{graphicx}
\usepackage{dcolumn}
\usepackage{bm}

\begin{document}

\preprint{APS/123-QED}

\title{Harmonic oscillator model for the helium atom}

\author{Martin Carlsen}
 \affiliation{Department of Applied Mathematics and Computer Science, Technical University of Denmark, Building 303B, DK-2800 Kongens Lyngby, Denmark}
 \email{ macar@dtu.dk}

\date{\today}

\begin{abstract}
A harmonic oscillator model in four dimensions is presented for the helium atom to estimate the distance to the inner and outer electron from the nucleus, the angle between electrons and the energy levels. The method is algebraic and is not based on the choice of correct trial wave function. Three harmonic oscillators and thus three quantum numbers are sufficient to describe the two-electron system. We derive a simple formula for the energy in the general case and in the special case of the Wannier Ridge. For a set of quantum numbers the distance to the electrons and the angle between the electrons are uniquely determined as the intersection between three surfaces. We show that the excited states converge either towards ionization thresholds or towards extreme parallel or antiparallel states and provide an estimate of the ground state energy.

\end{abstract}

\maketitle


\onecolumngrid
\section{Introduction}

The Bohr-Sommerfeld quantization method did not succeed to predict the ground state of helium and the three-body Coulomb potential has since then been one of the major challenges in quantum physics. The Bohr-Sommerfeld quantization failed since they assumed ad-hoc rules that later turned out not to be valid. They for instance assumed that the electron distance to the core is the same for the two electrons and that the quantum numbers are integers. The spectrum of helium can be divided into 1) bound states that consist of the ground state and discrete singly excited states (singlet and triplet states) 2) doubly excited states that often are autoionizing and 3) the continuum above the three-particle breakup threshold\cite{tanner2000theory}. The Hartree-Fock self-consistent-field method has shown to be particularly good at calculating bound states but it fails for doubly excited states where the electron-electron correlation is important (discovered by \cite{madden1963new} and discussed in \cite{fano1983correlations}). Numerical methods such as the complex rotation method based on a Hylleraas-type basis set or a Sturmian basis set using perimetric coordinates have been developed to study doubly excited states\cite{balslev1971spectral,ho1981complex,junker1982recent,junker1982recent,ho1983method,ho1985doubly,ho1986doubly,ho1991complex,ho1993doubly,lindroth1994calculation}. They have in addition been used to estimate the ground state energy with high accuracy\cite{salomonson1989solution}. An important question is why the single-particle picture of the Hartree-Fock method cannot be applied for doubly excited states. The molecular adiabatic approximation answers this question by considering the two-electron system as a $H_2^+$ molecule using the adiabatic distance as the interelectronic separation and the relative motion of the center of mass\cite{macek1968properties,feagin1986molecular,feagin1988molecular}. In this model the spatial wave function is separated into rotational, vibrational and molecular orbital wave functions, a more coherent picture than the single-particle picture. This model provides accurate estimations of the energy levels for doubly excited states. Classification methods to determine approximate quantum numbers of the two-electron system are primarily based on algebraic methods\cite{wulfman1968r,herrick1975comparison,herrick1983new,lin1984classification}. Finally, several semiclassical methods have been developed that consider the dynamics of a classical three-body potential\cite{gutzwiller1967phase,gutzwiller1971periodic,ezra1991semiclassical,wintgen1992semiclassical}. These are used to study chaos. 

The methods described above to determine the energy levels of a two-electron system  have in common that the choice of wave functions are crucial to the success of these methods (for the molecular adiabatic method for instance it leads to quasiseparability of the three-body Coulomb Schr\"odinger equation). The algebraic classification methods are used to determine the structure of the energy levels but do not by themselves predict the energy levels. This means that no succesful method to determine the bound and doubly excited energy levels of helium exists today that does not depend on the choice of spatial wave function. 
 
In this work, the energy levels of helium are determined using an algebraic method based on the harmonic oscillator in four dimensions\cite{moshinsky1969harmonic}. This leads to a system that consists of three oscillators and thus three quantum numbers $N_1$, $N_2$ and $N_3$. These three quantum numbers uniquely determine the energy $E$, the distance between the electrons and the nucleus $r_1$ and $r_2$ and the angle between the two electrons $\theta$. The ground state is found at $N_1=1$, $N_2=N_3=3/2$ with $E=-2.8827$a.u, $\tan\alpha=\frac{r_1}{r_2}=1.2635$, $r_2=1.0481${\AA} and $\cos\theta=-0.22725$. We analyze data for 1000 different combinations of the quantum numbers and derive a simple formula for the total energy. Finally, we consider the special case of the Wannier Ridge where $\tan\alpha=1$.

\section{Methods}
\subsection{The harmonic oscillator in 4 dimensions}
In this section, we derive the formulas to calculate $r=r_2$, $\cos\theta$, $\tan\alpha$ and $E$. The derivation is based on a second order expansion of the helium potential and the algebraic method for a simple harmonic oscillator generalized to $4$ dimensions. We derive the second order expansion in the next section.

 We consider the Hamiltonian
\begin{equation}\begin{aligned}
H&=\frac{1}{2m}\bm{p}^T\bm{p}+\frac{\hbar^2}{amr_{12}^3}\left(\frac{1}{2}\bm{x}^T\bm{C}^2\bm{x}-\bm{c}^T\bm{x}\right)+\text{constant},\label{H0}\\
\end{aligned}\end{equation}
where $\bm{x}$, $\bm{p}$, $\bm{c}$ and $\bm{C}^2$ are defined as
\begin{equation}\begin{aligned}
\bm{x}=\begin{pmatrix}x_1\\\vdots\\ x_4\end{pmatrix},\; \bm{p}=\begin{pmatrix}p_1\\\vdots\\p_4\end{pmatrix},\;\bm{c}=\begin{pmatrix}c_1\\\vdots\\c_4\end{pmatrix},\; \bm{C}^2=\begin{pmatrix}c_{11}&\hdots&c_{14}\\\vdots&\ddots&\vdots\\c_{41}&\hdots&c_{44}\end{pmatrix},
\end{aligned}\end{equation}
and $m$ is the mass of each electron, $\hbar$ is Planck's constant, $r_{12}$ is the distance between the electrons and $a$ is the Bohr radius defined by
\begin{equation}
a=\frac{4\pi\epsilon_0\hbar^2}{me^2}.
\end{equation}

In concordance with the standard algebraic method for simple harmonic oscillators we define the annihilation operator and creation operator as
\begin{equation}\begin{aligned}
\bm{\mu}&=\left(\frac{1}{4ar_{12}^3}\right)^{1/4}\left(\bm{C}\bm{x}+i\sqrt{\frac{ar^3}{\hbar^2}}\bm{p}\right)\\
\bm{\mu}^{\dag}&=\left(\frac{1}{4ar_{12}^3}\right)^{1/4}\left(\bm{C}\bm{x}-i\sqrt{\frac{ar^3}{\hbar^2}}\bm{p}\right).
\end{aligned}\end{equation}
Let $\bm{e}_i$ and $\lambda_i$ be the eigenvectors and eigenvalues of $\bm{C}^2$. We assume that $\lambda_i$ is nonnegative. The projections onto $\bm{e}_i$ are given by
\begin{equation}\begin{aligned}
x_i&=\bm{e}_i^T\bm{x}\\
p_i&=\bm{e}_i^T\bm{p}\\
\mu_i&=\bm{e}_i^T\bm{\mu}\\
\mu_i^{\dagger}&=\bm{e}_i^T\bm{\mu}^{\dagger}\\
c_i&=\bm{e}_i^T\bm{c}.
\end{aligned}\end{equation}
Clearly, $x_i$ and $p_i$ have the canonical commutation relation:
\begin{equation}\begin{aligned}
\left[x_i,p_i\right]=i\hbar.
\end{aligned}\end{equation}
When we add the creation and annihilation operator we find
\begin{equation}\begin{aligned}
\bm{C}\bm{x}=\frac{\left(ar_{12}^3\right)^{1/4}}{\sqrt{2}}\left(\bm{\mu}+\bm{\mu}^{\dag}\right),
\end{aligned}\end{equation}
so $x_i$ can be written as 
\begin{equation}\begin{aligned}
x_i=\frac{\left(ar_{12}^3\right)^{1/4}}{\sqrt{2}\sqrt{\lambda_i}}\left(\mu_i+\mu_i^{\dag}\right).
\end{aligned}\end{equation}
Using the canonical commutation relation, we find 
\begin{equation}\begin{aligned}
\left[\mu_i,\mu_i^{\dag}\right]&=\frac{-i\sqrt{\lambda_i}}{\hbar}\left[x_i,p_i\right]\\
&=\sqrt{\lambda_i}.
\end{aligned}\end{equation}
As a result, we have
\begin{equation}\begin{aligned}
\mu_i\mu_i^{\dag}&=\sqrt{\lambda_i} N_i+\frac{\sqrt{\lambda_i}}{2}\\
\mu_i^{\dag}\mu_i&=\sqrt{\lambda_i} N_i-\frac{\sqrt{\lambda_i}}{2}\\
\mu_i\mu_i^{\dag}+\mu_i^{\dag}\mu_i&=2\sqrt{\lambda_i}N_i.
\end{aligned}\end{equation}
It is easy to see that
\begin{equation}\begin{aligned}
\mu_i^{\dag}\mu_i&=\left(\frac{r_{12}}{a}\right)^{3/2}\frac{1}{2E_0}H_i-\frac{\sqrt{\lambda_i}}{2},
\end{aligned}\end{equation}
where
\begin{equation}\begin{aligned}
E_0=\frac{\hbar^2}{2ma^2},
\end{aligned}\end{equation}
so the total energy is given by
\begin{equation}\begin{aligned}
E&=\sum_i E_i\\
&=\sum_i2E_0\left(\frac{a}{r_{12}}\right)^{3/2}\sqrt{\lambda_i}N_i.
\end{aligned}\end{equation}
We made the assumption that $\lambda_i$ is nonnegative. This cannot be assumed in general. If $\lambda_i$ is nonpositive then $\sqrt{\lambda}_ii$ should be used instead of $\sqrt{\lambda}_i$ which ensures that $\sqrt{\lambda_i}N_ii$ is negative. For simplicity, we will assume that $N_i$ in this case is imaginary. It is also clear that $N_i$ is an integer. However, $N_i$ equal to $1/2$, $3/2$, $5/2$ is also a valid choice. We will thus assume that $N_i$ can be either half-integers or integers. 

To calculate $r$, $\cos\theta$ and $\tan\alpha$, we use
\begin{equation}\begin{aligned}
\left(\bm{c}^T\bm{e}_i\right)^2&=\frac{1}{2} \left(\bm{x}^T\bm{C}^2\bm{e}_i\right)^2,
\end{aligned}\end{equation}
or 
\begin{equation}\begin{aligned}
c_i^2&=\frac{1}{2}\lambda_i^2x_i^2,
\end{aligned}\end{equation}
which can be reduced to 
\begin{equation}\begin{aligned}
c_i^2&=\frac{1}{2}\lambda_i^2x_i^2\\
&= \frac{\lambda_i}{4}\sqrt{ar_{12}^3}\left(\mu_i+\mu_i^{\dag}\right)^2\\
&=\frac{\lambda_i^{3/2}}{2}\sqrt{ar_{12}^3}N_i.
\end{aligned}\end{equation}
We thus see that
\begin{equation}\begin{aligned}
r_{12}^3=\frac{4c_i^4}{N_i^2\lambda_i^3a}. \label{r}
\end{aligned}\end{equation}
This equation is very important. As $c_i$ is proportional to $r$ shown below this equation gives a formula for $r$ as a function of $\theta$ and $\alpha$ for each of the harmonic oscillators. Since the equation has to be satisfied for all of the oscillators it can be used to give an estimate of $r$, $\cos\theta$ and $\tan\alpha$ and thus the energy.

When using equation $\eqref{r}$ the energy can be written as
\begin{equation}\begin{aligned}
E&=\sum_iE_0\left(\frac{a\lambda_iN_i}{c_i}\right)^2=2E_0\frac{2 a}{r_{12}^3}\sum_i\frac{c_i^2}{ \lambda_i}.\label{H}
\end{aligned}\end{equation}
Finally, we remark that the method presented here easily can be generalized to any dimension.

\subsection{Second order expansion of the helium potential}

Let $r_1$ and $r_2$ be the distance between the electrons and the nucleus and let $r_{12}$ be the distance between the electrons. The helium potential is given by
\begin{equation}\begin{aligned}
V=\frac{\hbar^2}{am}\left(-\frac{2}{r_1}-\frac{2}{r_2}+\frac{1}{r_{12}}\right).
\end{aligned}\end{equation}
We define $\bm{x}_1$ and $\bm{x}_2$ as the coordinates of the electrons and $\bm{x}_0$ as the coordinates of the nucleus. The unit vectors $\bm{u}_1$, $\bm{v}_1$ and $\bm{w}_1$ are defined as
\begin{equation}\begin{aligned}
\bm{u}_1&=\frac{1}{r_1}\left(\bm{x}_1-\bm{x}_0\right),\;
\bm{v}_1=\frac{1}{r_2}\left(\bm{x}_2-\bm{x}_0\right),\;
\bm{w}_1=\frac{r_2}{r_{12}}\bm{v}_1-\frac{r_1}{r_{12}}\bm{u}_1.
\end{aligned}\end{equation}
Let $\left\{\bm{u}_1,\bm{u}_2,\bm{u}_3\right\}$, $\left\{\bm{v}_1,\bm{v}_2,\bm{v}_3\right\}$ and $\left\{\bm{w}_1,\bm{w}_2,\bm{w}_3\right\}$ be three orthonormal bases. Furthermore, let 
\begin{equation}\begin{aligned}
\tan\alpha=\frac{r_1}{r_2},\; r=r_2,
\end{aligned}\end{equation}
and
\begin{equation}\begin{aligned}
\bm{v}_1=\cos\theta\bm{u}_1+\sin\theta\bm{u}_{2}.
\end{aligned}\end{equation}
It follows that
\begin{equation}\begin{aligned}
r_{12}^2=r_1^2+r_2^2-2r_1r_2\cos\theta=r^2\left(1+\tan^2\alpha-2\tan\alpha\cos\theta\right).
\end{aligned}\end{equation}
In the three bases we define the operators:
\begin{equation}\begin{aligned}
\bm{P}_{1}&=\bm{u}_1\bm{u}_1^T, \; \bm{P}_{\perp}=I-\bm{P}_{1}\\
\bm{Q}_{1}&=\bm{v}_1\bm{v}_1^T, \; \bm{Q}_{\perp}=I-\bm{Q}_{1}\\
\bm{R}_{1}&=\bm{w}_1\bm{w}_1^T=\frac{1}{r_{12}^2}\left(r_1^2\bm{u}_1\bm{u}_1^T+r_{2}^2\bm{v}_1\bm{v}_1^T-r_{1}r_2\bm{u}_1\bm{v}_1^T-r_{1}r_2\bm{v}_1\bm{u}_1^T\right) \\&=\frac{1}{r_{12}^2}\left(\left(r_1-r_2\cos\theta\right)^2\bm{u}_1\bm{u}_1^T+r_2^2\sin^2\theta\bm{u}_2\bm{u}_2^T-\sin\theta\left(r_1r_2-r_2^2\cos\theta\right)\left(\bm{u}_1\bm{u}_2^T+\bm{u}_2\bm{u}_1^T\right)\right)\\
&=\frac{r^2}{r_{12}^2}\left(\left(\tan\alpha-\cos\theta\right)^2\bm{u}_1\bm{u}_1^T+\sin^2\theta\bm{u}_2\bm{u}_2^T-\sin\theta\left(\tan\alpha-\cos\theta\right)\left(\bm{u}_1\bm{u}_2^T+\bm{u}_2\bm{u}_1^T\right)\right)\\
 \bm{R}_{\perp}&=I-\bm{R}_{1},
\end{aligned}\end{equation}
such that $\bm{P}_{1}$ and $\bm{P}_{\perp}$ in the basis $\left\{\bm{u}_1,\bm{u}_2,\bm{u}_3\right\}$ have the matrix representations
\begin{equation}
\bm{P}_{1}=\begin{bmatrix}1 &0&0\\0&0&0\\0&0&0\end{bmatrix},\; \bm{P}_{\perp}=\begin{bmatrix}0 &0&0\\0&1&0\\0&0&1\end{bmatrix},
\end{equation}
and similarly for $\bm{Q}_{1}$ and $\bm{Q}_{\perp}$ in the basis $\left\{\bm{v}_1,\bm{v}_2,\bm{v}_3\right\}$ and $\bm{R}_{1}$ and $\bm{R}_{\perp}$ in the basis $\left\{\bm{w}_1,\bm{w}_2,\bm{w}_3\right\}$.

The second order perturbation of $V$ is given by\cite{carlsen2014using}
\begin{equation}\begin{aligned}
\begin{split}
&\frac{\hbar^2}{am}\left(\frac{2}{r_1^2}\bm{x}^T\bm{u}_1\frac{2}{r_2^2}\bm{y}^T\bm{v}_1-\frac{1}{r_{12}^2}\left(\bm{x}-\bm{y}\right)^T\bm{w}_1\right.\\
&\left.-\frac{1}{r_1^3}\bm{x}^T\left(2\bm{P}_1-\bm{P}_{\perp}\right)\bm{x}-\frac{1}{r_2^3}\bm{y}^T\left(2\bm{Q}_1-\bm{Q}_{\perp}\right)\bm{y}
+\frac{1}{2r_{12}^3}\left(\bm{x}-\bm{y}\right)^T\left(2\bm{R}_1-\bm{R}_{\perp}\right)\left(\bm{x}-\bm{y}\right)\right),
\end{split}
\end{aligned}\end{equation}
where I have subtracted the zeroth order terms. This motivates us to consider the following potential
\begin{equation}\begin{aligned}
\begin{split}
&\frac{\hbar^2}{am}\left(\frac{2}{r_1^2}\bm{x}^T\bm{u}_1-\frac{2}{r_2^2}\bm{y}^T\bm{v}_1-\frac{1}{r_{12}^2}\left(\bm{x}-\bm{y}\right)^T\bm{w}_1-\frac{1}{r_1^3}\bm{x}^T2\bm{P}_1\bm{x}-\frac{1}{r_2^3}\bm{y}^T2\bm{Q}_1\bm{y}
+\frac{1}{2r_{12}^3}\left(\bm{x}-\bm{y}\right)^T2\bm{R}_1\left(\bm{x}-\bm{y}\right)\right).
\end{split}
\end{aligned}\end{equation}
There are two important changes: 1) We only consider terms involving $\bm{u}_1$, $\bm{v}_1$ and $\bm{w}_1$ and 2) We consider $\bm{x}+\bm{y}$ instead of $\bm{x}-\bm{y}$ and let $\bm{y}\rightarrow -\bm{y}$.

This is an atomic system in four dimensions. Define $\bm{x}=x_1\bm{u}_1+x_1\bm{u}_2$ and $\bm{y}=y_1\bm{u}_1+y_1\bm{u}_2$. The $4\times 4$ matrix $\bm{C}^2$ in equation \eqref{H0} is given by
\begin{equation}\begin{aligned}
\bm{C}^2&=-4r_{12}^3\begin{pmatrix}
\frac{1}{r_1^3}&0&0&0\\0&0&0&0\\
0&0&\frac{1}{r_2^3}\cos^2\theta&\frac{1}{r_2^3}\cos\theta\sin\theta\\
0&0&\frac{ 1}{r_2^3}\cos\theta\sin\theta&\frac{1}{r_2^3}\sin^2\theta
\end{pmatrix}\\
&-\frac{2}{r_{12}^2}\begin{pmatrix}
-\left(r_1-r_2\cos\theta\right)^2&\sin\theta\left(r_1r_2-r_2^2\cos\theta\right)&\left(r_1-r_2\cos\theta\right)^2&-\sin\theta\left(r_1r_2-r_2^2\cos\theta\right)\\
\sin\theta\left(r_1r_2-r_2^2\cos\theta\right)&-r_2^2\sin^2\theta&-\sin\theta\left(r_1r_2-r_2^2\cos\theta\right)&r_2^2\sin^2\theta\\
\left(r_1-r_2\cos\theta\right)^2&-\sin\theta\left(r_1r_2-r_2^2\cos\theta\right)&-\left(r_1-r_2\cos\theta\right)^2&\sin\theta\left(r_1r_2-r_2^2\cos\theta\right)\\
-\sin\theta\left(r_1r_2-r_2^2\cos\theta\right)&r_2^2\sin^2\theta&\sin\theta\left(r_1r_2-r_2^2\cos\theta\right)&-r_2^2\sin^2\theta
\end{pmatrix},
\end{aligned}\end{equation}
and the vector $\bm{c}$ is given by
\begin{equation}\begin{aligned}
\bm{c}&=\begin{pmatrix}\frac{2r_{12}^3}{r_1^2}+r_{1}-r_2\cos\theta\\-r_2\sin\theta\\-\frac{2r_{12}^3}{r_2^2}\cos\theta-r_1+r_2\cos\theta\\-\frac{2r_{12}^3}{r_2^2}\sin\theta+r_2\sin\theta\end{pmatrix}.
\end{aligned}\end{equation}
They can be written as
\begin{equation}\begin{aligned}
\bm{C}^2&=-4\frac{r_{12}^3}{r^3}\begin{pmatrix}\tan^{-3}\alpha&0&0&0\\0&0&0&0\\0&0&\cos^2\theta&\cos\theta\sin\theta\\0&0&\cos\theta\sin\theta&\sin^2\theta\end{pmatrix}\\
&-\frac{2r^2}{r_{12}^2}\begin{pmatrix}
-\left(\tan\alpha-\cos\theta\right)^2&\sin\theta\left(\tan\alpha-\cos\theta\right)&\left(\tan\alpha-\cos\theta\right)^2&-\sin\theta\left(\tan\alpha-\cos\theta\right)\\
\sin\theta\left(\tan\alpha-\cos\theta\right)&-\sin^2\theta&-\sin\theta\left(\tan\alpha-\cos\theta\right)&\sin^2\theta\\
\left(\tan\alpha-\cos\theta\right)^2&-\sin\theta\left(\tan\alpha-\cos\theta\right)&-\left(\tan\alpha-\cos\theta\right)^2&\sin\theta\left(\tan\alpha-\cos\theta\right)\\
-\sin\theta\left(\tan\alpha-\cos\theta\right)&\sin^2\theta&\sin\theta\left(\tan\alpha-\cos\theta\right)&-\sin^2\theta^2
\end{pmatrix},
\end{aligned}\end{equation}
and
\begin{equation}\begin{aligned}
\bm{c}&=r\begin{pmatrix}2\frac{r_{12}^3}{r^3}\tan^{-2}\alpha+\tan\alpha-\cos\theta\\-\sin\theta\\-2\frac{r_{12}^3}{r^3}\cos\theta-\tan\alpha+\cos\theta\\-2\frac{r_{12}^3}{r^3}\sin\theta+\sin\theta\end{pmatrix}.
\end{aligned}\end{equation}
The eigenvectors and eigenvalues of $\bm{C}^2$ can be found numerically so the expressions for $\bm{C}^2$ and $\bm{c}$ as a function of $\theta$ and $\alpha$ are enough to use the formulas derived in the preceding section. It is, however, useful to derive the analytical expression for $\bm{e}_i$ and $\lambda_i$ as it can be used to prove a formula for the energy. 

\subsection{Derivation of $\bm{e}_i$, $\lambda_i$ and $c_i$}
The eigenvalues and eigenvectors of $\bm{C}^2$ can be calculated using the equations 
{\small
\begin{equation}\begin{aligned}
-\frac{r_{12}^2}{2r^2}\lambda a&=\left(\frac{2r_{12}^5}{r^5}\tan^{-3}\alpha-\left(\tan\alpha-\cos\theta\right)^2\right)a +\sin\theta\left(\tan\alpha-\cos\theta\right)b+\left(\tan\alpha-\cos\theta\right)^2c-\sin\theta\left(\tan\alpha-\cos\theta\right)d\\
-\frac{r_{12}^2}{2r^2}\lambda b&=\sin\theta\left(\tan\alpha-\cos\theta\right)a-\sin^2\theta b-\sin\theta\left(\tan\alpha-\cos\theta\right)c+\sin^2\theta d\\
-\frac{r_{12}^2}{2r^2}\lambda c&=\left(\tan\alpha-\cos\theta\right)^2a-\sin\theta\left(\tan\alpha-\cos\theta\right)b+\left(\frac{2r_{12}^5}{r^5}\cos^2\theta-\left(\tan\alpha-\cos\theta\right)\right)c+\left(\frac{2r_{12}^5}{r^5}\cos\theta\sin\theta+\sin\theta\left(\tan\alpha-\cos\theta\right)\right)d\\
-\frac{r_{12}^2}{2r^2}\lambda d&=-\sin\theta\left(\tan\alpha-\cos\theta\right)a+\sin^2\theta b+\left(\frac{2r_{12}^5}{r^5}\cos\theta\sin\theta+\sin\theta\left(\tan\alpha-\cos\theta\right)\right)c+\left(\frac{2r_{12}^5}{r^5}\sin^2\theta-\sin^2\theta\right)d.
\end{aligned}\end{equation}}
From these four equations, we deduce 
\begin{equation}\begin{aligned}
-\frac{r_{12}^2}{2r^2}\lambda\left(\sin\theta a+\left(\tan\alpha-\cos\theta\right)b\right)&=\frac{2r_{12}^5}{r^5}\tan^{-3}\alpha\sin\theta a   \\
-\frac{r_{12}^2}{2r^2}\lambda\left(\sin\theta c-\left(\tan\alpha-\cos\theta\right)b\right)&=\frac{2r_{12}^5}{r^5}\left(\cos^2\theta c +\cos\theta\sin\theta d\right)\sin\theta\\
-\frac{r_{12}^2}{2r^2}\lambda\left(\sin\theta a -\left(\tan\alpha-\cos\theta\right)d\right)&=\frac{2r_{12}^5}{r^5}\left(a\sin\theta\tan^{-3}\alpha-\cos\theta\sin\theta\left(\tan\alpha-\cos\theta\right)c-\sin^2\theta\left(\tan\alpha-\cos\theta\right)d\right)\\
-\frac{r_{12}^2}{2r^2}\lambda \left(a+c\right)&=\frac{2r_{12}^5}{r^5}\left(a\tan^{-3}\alpha+\cos^2\theta c + \cos\theta\sin\theta d\right)\\
-\frac{r_{12}^2}{2r^2}\lambda \left(b+d\right)&=\frac{2r_{12}^5}{r^5}\left(\cos\theta\sin\theta c + \sin^2\theta d\right)\\
-\frac{r_{12}^2}{2r^2}\lambda \left(\sin\theta \left(a+c\right)-\cos\theta\left(d+b\right)\right)&=\frac{2r_{12}^5}{r^5}\tan^{-3}\alpha\sin\theta a.
\end{aligned}\end{equation}
First we find that $\lambda=0$ is an eigenvalue with $a=0$, $b=1$, $c=-\frac{\sin\theta}{\tan\alpha}$ and $d=\frac{-\cos\theta}{\sin\theta} c=\frac{\cos\theta}{\tan\alpha}$. Next, we can without loss of generality assume that $a=1$. When combining the equations we find the relation
\begin{equation}\begin{aligned}
\sin\theta c - \cos\theta d &=b\tan\alpha.
\end{aligned}\end{equation}
Furthermore, we obtain
\begin{equation}\begin{aligned}
\frac{1}{\tan\alpha\sin\theta + \left(\tan\alpha-\cos\theta \right)\left(\sin\theta c - \cos\theta d \right)}=\frac{\left(\cos\theta c+\sin\theta d\right)\tan^3\alpha}{\sin\theta c+\left(\tan\alpha-\cos\theta\right)d}.
\end{aligned}\end{equation}
This equation can be reduced to the quadratic equation 
\begin{equation}\begin{aligned}
&\left(1-\tan^4\alpha\cos\theta\right)\sin\theta c+\left( \left(\tan\alpha-\cos\theta\right)-\tan^4\alpha\sin^2\theta\right)d+\left(\tan\alpha-\cos\theta\right)\tan^3\alpha\cos\theta\sin\theta\left(d^2-c^2\right)\\
&+\left(\tan\alpha-\cos\theta\right)\tan^3\alpha\left(\cos^2\theta-\sin^2\theta\right)cd=0,
\end{aligned}\end{equation}
which can be written as
\begin{equation}\begin{aligned}
&\sin\theta c - \cos\theta d -\tan^4\alpha \sin\theta \left(\cos\theta c + \sin \theta d\right)+\tan\alpha d-\left(\tan\alpha-\cos\theta\right)\tan^3\alpha\left(\sin\theta c-\cos\theta d\right)\left(\cos\theta c + \sin \theta d\right)\\
&=\tan\alpha\sin\theta\left(1-\tan^3\alpha\right)\gamma_1+\left(1-\tan\alpha\cos\theta\right)\gamma_2-\left(\tan\alpha-\cos\theta\right)\tan^3\alpha\gamma_1\gamma_2=0,
\end{aligned}\end{equation}
or
\begin{equation}\begin{aligned}
-\left(\gamma_1-\frac{1-\tan\alpha\cos\theta}{\left(\tan\alpha-\cos\theta\right)\tan^3\alpha}\right)\left(\gamma_2-\frac{\tan\alpha\sin\theta\left(1-\tan^3\alpha\right)}{\left(\tan\alpha-\cos\theta\right)\tan^3\alpha}\right)+\frac{\left(1-\tan\alpha\cos\theta\right)\left(1-\tan^3\alpha\right)\sin\theta}{\left(\tan\alpha-\cos\theta\right)^2\tan^5\alpha}=0, \label{gamma12}
\end{aligned}\end{equation}
where $\gamma_1$ and $\gamma_2$ are defined as
\begin{equation}\begin{aligned}
\gamma_1&=\cos\theta c+\sin\theta d\\
\gamma_2&=\sin\theta c-\cos\theta d,
\end{aligned}\end{equation}
and $c$ and $d$ are given by
\begin{equation}\begin{aligned}
c&=\cos\theta \gamma_1+\sin\theta \gamma_2\\
d&=\sin\theta \gamma_1-\cos\theta \gamma_2.
\end{aligned}\end{equation}
For $\gamma_1$ we have 
\begin{equation}\begin{aligned}
\gamma_1=\frac{\left(1-\tan^3\alpha\right)\left(1-\tan\alpha\cos\theta\right)\sin\theta}{\left(\tan\alpha-\cos\theta\right)^2\tan^5\alpha\gamma_2-\tan^3\alpha\sin\theta\left(1-\tan^3\alpha\right)\left(\tan\alpha-\cos\theta\right)}+\frac{1-\tan\alpha\cos\theta}{\left(\tan\alpha-\cos\theta\right)\tan^3\alpha}.
\end{aligned}\end{equation}
The following equality also holds:
\begin{equation}\begin{aligned}
\sin\theta\left(\tan\alpha-\cos\theta\right)-\sin^2\theta\frac{\gamma_2}{\tan\alpha}-\sin\theta\tan\alpha\left(\cos\theta\gamma_1+\sin\theta\gamma_2\right)+\sin\theta\gamma_1=\frac{2\frac{r_{12}^5}{r^5}\tan^{-3}\alpha\sin\theta\frac{\gamma_2}{\tan\alpha}}{\sin\theta+\left(\tan\alpha-\cos\theta\right)\frac{\gamma_2}{\tan\alpha}},
\end{aligned}\end{equation}
and it leads to the quadratic equation
\begin{equation}\begin{aligned}
&\sin\theta\left(\tan\alpha-\cos\theta\right)+\sin\theta\left(1-\tan\alpha\cos\theta\right)\gamma_1+\left(-\frac{2r_{12}^5}{r^5}\tan^{-4}\alpha+\frac{\left(\tan\alpha-\cos\theta\right)^2}{\tan\alpha}-\frac{\sin^2\theta}{\tan\alpha}-\sin^2\theta\tan\alpha\right)\gamma_2\\
&+\left(1-\tan\alpha\cos\theta\right)\frac{\left(\tan\alpha-\cos\theta\right)}{\tan\alpha}\gamma_1\gamma_2-\sin\theta\left(\frac{1}{\tan^2\alpha}+1\right)\left(\tan\alpha-\cos\theta\right)\gamma_2^2=0.
\end{aligned}\end{equation}
Using the equation above for $\gamma_1\gamma_2$ we get
\begin{equation}\begin{aligned}
&\sin\theta\left(\tan\alpha-\cos\theta\right)+\sin\theta\tan^{-3}\alpha\left(1-\tan\alpha\cos\theta\right)\gamma_1\\
&+\left(-\frac{2r_{12}^5}{r^5}\tan^{-4}\alpha+\frac{\left(\tan\alpha-\cos\theta\right)^2}{\tan\alpha}-\frac{\sin^2\theta}{\tan\alpha}-\sin^2\theta\tan\alpha+\tan^{-4}\alpha\left(1-\tan\alpha\cos\theta\right)^2\right)\gamma_2\\
&-\sin\theta\left(\frac{1}{\tan^2\alpha}+1\right)\left(\tan\alpha-\cos\theta\right)\gamma_2^2=0,
\end{aligned}\end{equation}
and inserting the equation for $\gamma_1$ we finally obtain the cubic equation
{\small
\begin{equation}\begin{aligned}
&-\sin^2\theta\left(1-\tan^3\alpha\right)\tan^{3}\alpha\left(\tan\alpha-\cos\theta\right)^2\\
&+\left(\sin\theta\tan^{-1}\alpha\left(1-\tan\alpha\cos\theta\right)^2+\sin\theta\tan^5\alpha\left(\tan\alpha-\cos\theta\right)^2\right)\left(\tan\alpha-\cos\theta\right)\gamma_2\\
&-\tan^3\alpha\sin\theta\left(1-\tan^3\alpha\right)\left(-\frac{2r_{12}^5}{r^5}\tan^{-4}\alpha+\frac{\left(\tan\alpha-\cos\theta\right)^2}{\tan\alpha}-\frac{\sin^2\theta}{\tan\alpha}-\sin^2\theta\tan\alpha+\tan^{-4}\alpha\left(1-\tan\alpha\cos\theta\right)^2\right)\left(\tan\alpha-\cos\theta\right)\gamma_2\\
&+\tan^3\alpha\sin^2\theta\left(1-\tan^3\alpha\right)\left(\frac{1}{\tan^2\alpha}+1\right)\left(\tan\alpha-\cos\theta\right)^2\gamma_2^2\\
&+\tan^5\alpha\left(-\frac{2r_{12}^5}{r^5}\tan^{-4}\alpha+\frac{\left(\tan\alpha-\cos\theta\right)^2}{\tan\alpha}-\frac{\sin^2\theta}{\tan\alpha}-\sin^2\theta\tan\alpha+\tan^{-4}\alpha\left(1-\tan\alpha\cos\theta\right)^2\right)\left(\tan\alpha-\cos\theta\right)^2\gamma_2^2\\
&-\tan^5\alpha\sin\theta\left(\frac{1}{\tan^2\alpha}+1\right)\left(\tan\alpha-\cos\theta\right)^3\gamma_2^3=0.\label{cubeq}
\end{aligned}\end{equation}}
We denote the solutions to this equation $\gamma_{2,1}$, $\gamma_{2,2}$ and $\gamma_{2,3}$. $\gamma_{1,1}$, $\gamma_{1,2}$ and $\gamma_{1,3}$ can be found by calculating $\gamma_1$. The eigenvectors of $\bm{C}^2$ are thus
\begin{equation}\begin{aligned}
\bm{e}_1&=\frac{1}{\sqrt{1+\gamma_{2,1}^2\left(1+\tan^{-2}\alpha\right)+\gamma_{1,1}^2 }}\begin{pmatrix}1\\\tan^{-1}\alpha\gamma_{2,1}\\\cos\theta\gamma_{1,1}+\sin\theta\gamma_{2,1}\\\sin\theta\gamma_{1,1}-\cos\theta\gamma_{2,1} \end{pmatrix},\\
\bm{e}_2&=\frac{1}{\sqrt{1+\gamma_{2,2}^2\left(1+\tan^{-2}\alpha\right)+\gamma_{1,2}^2}}\begin{pmatrix}1\\\tan^{-1}\alpha\gamma_{2,2}\\\cos\theta\gamma_{1,2}+\sin\theta\gamma_{2,2}\\\sin\theta\gamma_{1,2}-\cos\theta\gamma_{2,2}  \end{pmatrix},\\
\bm{e}_3&=\frac{1}{\sqrt{1+\gamma_{2,3}^2\left(1+\tan^{-2}\alpha\right)+\gamma_{1,3}^2}}\begin{pmatrix}1\\\tan^{-1}\alpha\gamma_{2,3}\\\cos\theta\gamma_{1,3}+\sin\theta\gamma_{2,3}\\\sin\theta\gamma_{1,3}-\cos\theta\gamma_{2,3}  \end{pmatrix},\\
\bm{e}_4&=\frac{1}{\sqrt{1+\tan^{-2}\alpha}}\begin{pmatrix}0\\1\\-\sin\theta\tan^{-1}\alpha\\\cos\theta\tan^{-1}\alpha \end{pmatrix}.
\end{aligned}\end{equation}
The eigenvalues of $\bm{C}^2$ are 
\begin{equation}\begin{aligned}
\lambda_1&=-\frac{2r^2}{r_{12}^2\gamma_{2,1}}\sin\theta\tan\alpha \xi_1\\
\lambda_2&=-\frac{2r^2}{r_{12}^2\gamma_{2,2}}\sin\theta\tan\alpha \xi_2\\
\lambda_3&=-\frac{2r^2}{r_{12}^2\gamma_{2,3}}\sin\theta\tan\alpha \xi_3\\
\lambda_4&=0,
\end{aligned}\end{equation}
where $\xi_i$ are defined as
\begin{equation}\begin{aligned}
\xi_i=\tan\alpha-\cos\theta+\left(1-\tan\alpha\cos\theta\right)\gamma_{1,i}-\sin\theta\tan^{-1}\alpha\left(1+\tan^2\alpha\right)\gamma_{2,i}.
\end{aligned}\end{equation}
Using the expressions for $\bm{e}_i$ we find that $c_i=\bm{e}_i^T\bm{c}_i$ are given by
\begin{equation}\begin{aligned}
c_1&=r\frac{2\frac{r_{12}^3}{r^3}\left(\tan^{-2}\alpha-\gamma_{1,1}\right)+\xi_1}{\sqrt{1+\gamma_{2,1}^2\left(1+\tan^{-2}\alpha\right)+\gamma_{1,1}^2 }}\\
c_2&=r\frac{2\frac{r_{12}^3}{r^3}\left(\tan^{-2}\alpha-\gamma_{1,2}\right)+\xi_2}{\sqrt{1+\gamma_{2,2}^2\left(1+\tan^{-2}\alpha\right)+\gamma_{1,2}^2 }}\\
c_3&=r\frac{2\frac{r_{12}^3}{r^3}\left(\tan^{-2}\alpha-\gamma_{1,3}\right)+\xi_3}{\sqrt{1+\gamma_{2,3}^2\left(1+\tan^{-2}\alpha\right)+\gamma_{1,3}^2 }}\\
c_4&=0.
\end{aligned}\end{equation}

\subsection{Formula for $E$}
In this section we prove that the energy is given by the expression:
\begin{equation}\begin{aligned}
\begin{split}
E=2E_0\frac{2a}{r}\left(-1-\tan^{-1}\alpha+\frac{r}{2r_{12}}\right).\label{Eh}
\end{split}
\end{aligned}\end{equation}
First, the orthogonality of the eigenvectors implies 
\begin{equation}\begin{aligned}
1+\gamma_{1,1}\gamma_{1,2}+\left(1+\tan^{-2}\alpha\right)\gamma_{2,1}\gamma_{2,2}=0\\
1+\gamma_{1,1}\gamma_{1,3}+\left(1+\tan^{-2}\alpha\right)\gamma_{2,1}\gamma_{2,3}=0\\
1+\gamma_{1,2}\gamma_{1,3}+\left(1+\tan^{-2}\alpha\right)\gamma_{2,2}\gamma_{2,3}=0. \label{ortheq}
\end{aligned}\end{equation}
Using the three equations we deduce the following useful relations
\begin{equation}\begin{aligned}
\sum_i\frac{1}{1+\gamma_{1,i}^2+\left(1+\tan^2\alpha\right)\gamma_{2,i}^2}&=1,\; \sum_i\frac{\gamma_{1,i}}{1+\gamma_{1,i}^2+\left(1+\tan^2\alpha\right)\gamma_{2,i}^2}=0,\; \sum_i\frac{\gamma_{2,i}}{1+\gamma_{1,i}^2+\left(1+\tan^2\alpha\right)\gamma_{2,i}^2}=0\\
\sum_i\frac{\gamma_{1,i}^2}{1+\gamma_{1,i}^2+\left(1+\tan^2\alpha\right)\gamma_{2,i}^2}&=1,\; \sum_i\frac{\gamma_{2,i}^2}{1+\gamma_{1,i}^2+\left(1+\tan^2\alpha\right)\gamma_{2,i}^2}=\frac{\tan^2\alpha}{1+\tan^2\alpha},\; \sum_i\frac{\gamma_{1,i}\gamma_{2,i}}{1+\gamma_{1,i}^2+\left(1+\tan^2\alpha\right)\gamma_{2,i}^2}=0\\
\sum_i\frac{1}{\xi_i}\frac{\gamma_{1,i}\gamma_{2,i}}{1+\gamma_{1,i}^2+\left(1+\tan^2\alpha\right)\gamma_{2,i}^2}&=0,\; \sum_i\frac{1}{\xi_i}\frac{\gamma_{2,i}}{1+\gamma_{1,i}^2+\left(1+\tan^2\alpha\right)\gamma_{2,i}^2}=\frac{\sin\theta\tan^4\alpha r^5}{2r_{12}^5},\\
\sum_i\frac{1}{\xi_i}\frac{\gamma_{1,i}^2\gamma_{2,i}}{1+\gamma_{1,i}^2+\left(1+\tan^2\alpha\right)\gamma_{2,i}^2}&=\frac{\sin\theta\tan\alpha r^5}{2r_{12}^5}.\label{rel}
\end{aligned}\end{equation}
The first relation is proven in the Appendix A. Given the above relations we can readily prove the formula for the energy. It is given by
\begin{equation}\begin{aligned}
\begin{split}
E&=2E_0\frac{2a}{r_{12}^3}\sum_i\frac{c_i^2}{\lambda_i}\\
&=-2E_0\frac{a}{r}\frac{r}{r_{12}}\frac{1}{\sin\theta\tan\alpha}\sum_i\frac{\gamma_{2,i}\left(2\frac{r_{12}^3}{r^3}\left(\tan^{-2}\alpha-\gamma_{1,i}\right)+\xi_i\right)^2}{\xi_i\left(1+\gamma_{1,i}^2+\gamma_{2,i}^2\left(1+\tan^{-2}\alpha\right)\right)}\\
&=-2E_0\frac{a}{r}\frac{r}{r_{12}}\frac{1}{\sin\theta\tan\alpha}\sum_i\gamma_{2,i}\frac{\frac{4\frac{r_{12}^6}{r^6}\left(\tan^{-2}\alpha-\gamma_{1,i}\right)^2}{\xi_i}+4\frac{r_{12}^3}{r^3}\left(\tan^{-2}\alpha-\gamma_{1,i}\right)+\xi_i}{1+\gamma_{1,i}^2+\gamma_{2,i}^2\left(1+\tan^{-2}\alpha\right)}.
\end{split}
\end{aligned}\end{equation}
Several terms vanish due to the relations mentioned above. The non-vanishing terms are
\begin{equation}\begin{aligned}
\begin{split}
E=-2E_0\frac{a}{r}\frac{r}{r_{12}}\frac{1}{\sin\theta\tan\alpha}\sum_i\frac{\frac{4\frac{r_{12}^6}{r^6}\left(\tan^{-4}\alpha\gamma_{2,i}+\gamma_{1,i}^2\gamma_{2,i}\right)}{\xi_i}-\sin\theta\tan^{-1}\alpha\left(1+\tan^2\alpha\right)\gamma_{2,i}^2}{1+\gamma_{1,i}^2+\gamma_{2,i}^2\left(1+\tan^{-2}\alpha\right)},
\end{split}
\end{aligned}\end{equation}
so we finally obtain the expression for the energy of the helium atom
\begin{equation}\begin{aligned}
\begin{split}
E=2E_0\frac{2a}{r}\left(-1-\tan^{-1}\alpha+\frac{r}{2r_{12}}\right).
\end{split}
\end{aligned}\end{equation}
The derivation is not valid when $\tan\alpha=1$. The formula for the energy in this special case is proven in Appendix B.

\subsection{Numerical estimation of $r$, $\tan\alpha$, $\cos\theta$ and $E$}
\begin{figure}[t!]
\centering
\includegraphics[width=0.5\textwidth,keepaspectratio=true]{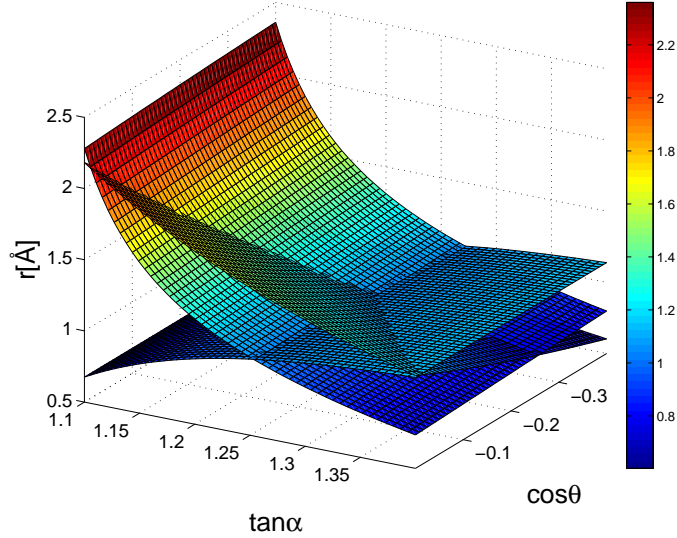}
\caption{Showing the three surfaces for $N_1=1$, $N_2=N_3=3/2$. Equation \eqref{r} is satisfied for all of the three oscillators when the three surfaces intersect.}\label{fig3}
\end{figure}
Equation \eqref{r} viewed as the function $r(\theta,\alpha,N_i)$  determines $r$ for each oscillator. To evaluate $r$ requires a calculation of $\bm{e}_i$, $\lambda_i$ and $c_i$ that were found by calculating the three analytical solutions to equation \eqref{cubeq}. For a given $N_i$, $r$ is a surface as a function $\theta$ and $\alpha$. It is clear that  $r$, $\theta$ and $\alpha$ must be uniquely determined. Hence, $r$, $\theta$ and $\alpha$ can be found by calculating the intersection points between the three surfaces. Observations suggest that only one intersection point exists allthough no proof has been given in this work. Given $N_1$, $N_2$ and $N_3$,  $r$, $\theta$ and $\alpha$ can thus be estimated by calculating the unique intersection point where equation \eqref{r} is satisfied for all of the three oscillators. This is illustrated in Figure \ref{fig3}. Let $s_1$, $s_2$ and $s_3$ be $r$ for each of the three oscillators. The intersection point is estimated by calculating
\begin{equation}
\text{residual}=\min_{\theta,\alpha}\left(|s_1-s_2|+|s_1-s_3| + |s_2-s_3|\right)\label{res}
\end{equation}
using a fine grid of $1000\times 1000$ where the range of $\cos\theta$ is from $-0.99999$ to $0.99999$ and the range of $\tan\alpha$ is from $1.00001$ to $10$. The method is then iterated by decreasing the range of $\cos\theta$ and $\tan\alpha$ to be close to the estimated intersection point. For the estimated values of $r$, $\theta$ and $\alpha$ the energy can be calculated either by use of the analytical formula derived above or by equation \eqref{H}.

\section{Results}
The observables $r$, $\tan\alpha$, $\cos\theta$ and $E$ were calculated using the intersection method described in the preceding section for different values of $N_1$, $N_2$ and $N_3$ under the assumption that the quantum numbers are either half-integers or integers. The estimations of $r$, $\tan\alpha$, $\cos\theta$ and $E$ and the residuals given by equation \eqref{res} for 1000 different combinations of $N_1$, $N_2$ and $N_3$ with a range from $0.5$ to $5$ were tabulated in the supporting material. 
\begin{figure}[t!]
\centering
\subfloat[]{\label{fig2a}%
\includegraphics[width=0.48\textwidth,keepaspectratio=true]{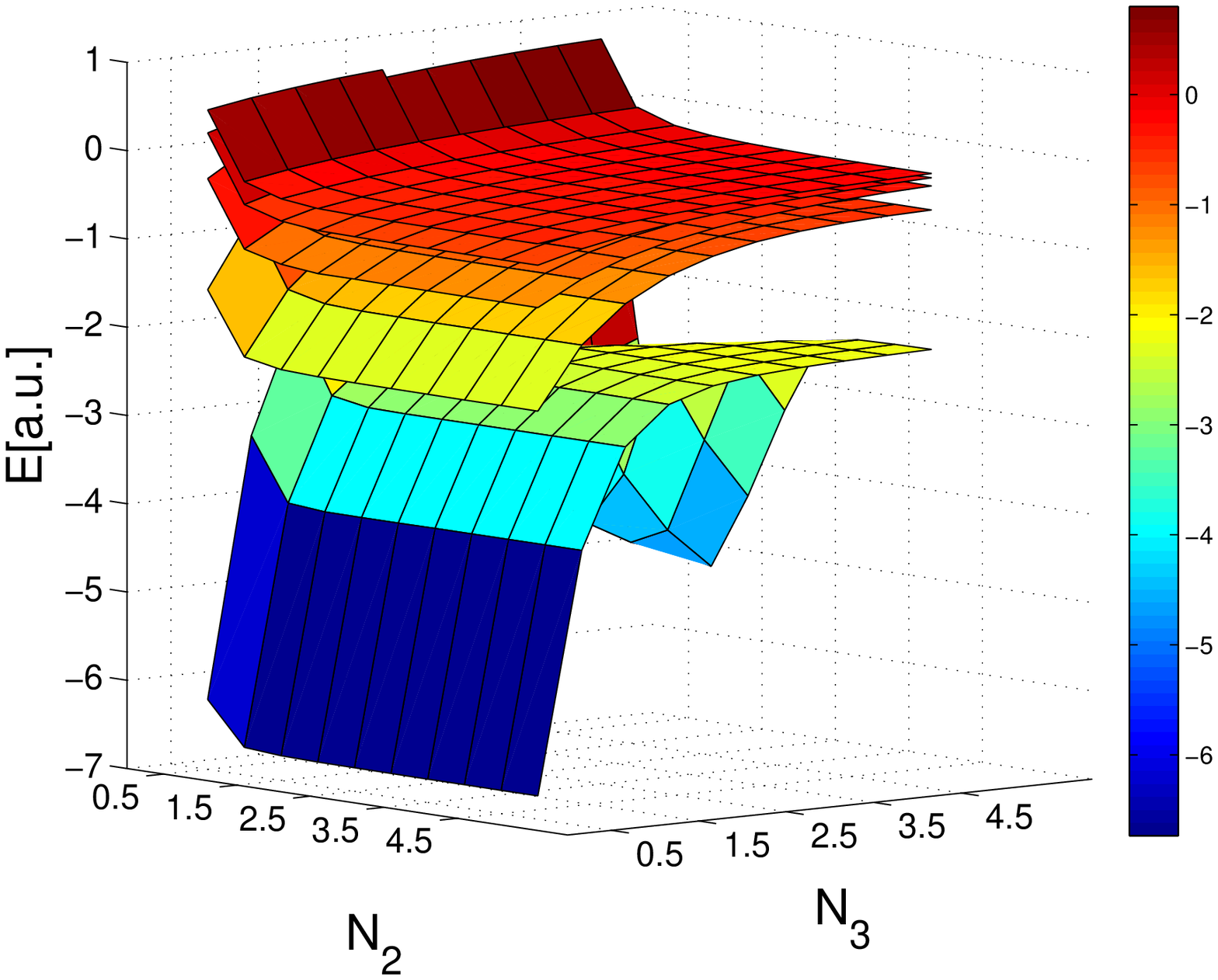}
}
\subfloat[]{\label{fig2b}%
\includegraphics[width=0.48\textwidth,keepaspectratio=true]{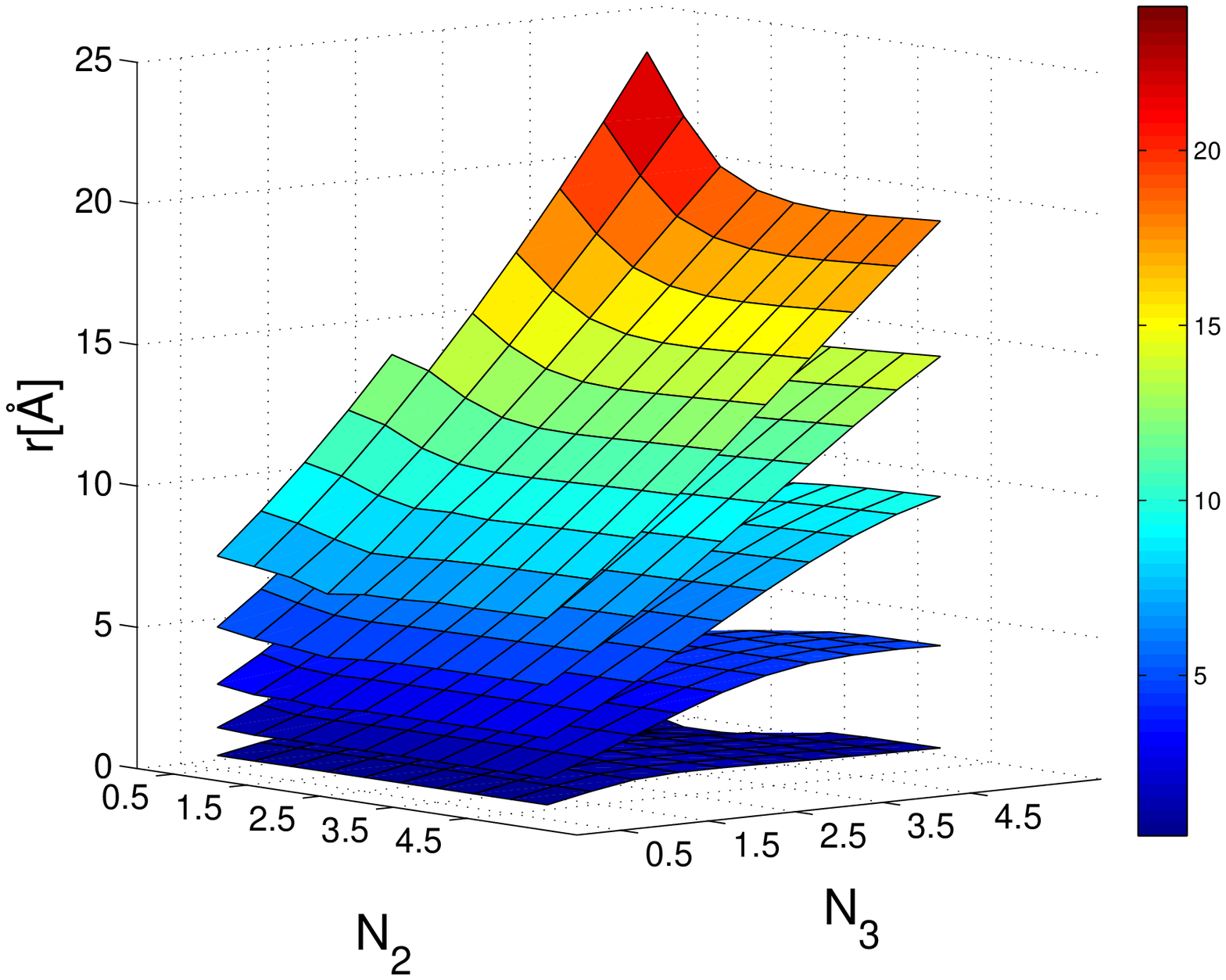}
}
\caption{For $N_1=1,2,3,4$ and $5$ we plot the energy (left) and the distance to the inner electron (right) for different values of $N_2$ and $N_3$. The surface plot of $N_1=1$ is lowest. The surface plot of $N_1=2$ is between $N_1=1$ and $N_1=3$ and so forth. The data can also be found in the supporting material. }
\label{fig2}
\end{figure}

\begin{figure}[t!]
\centering
\subfloat[]{\label{fig1a}%
\includegraphics[width=0.48\textwidth,keepaspectratio=true]{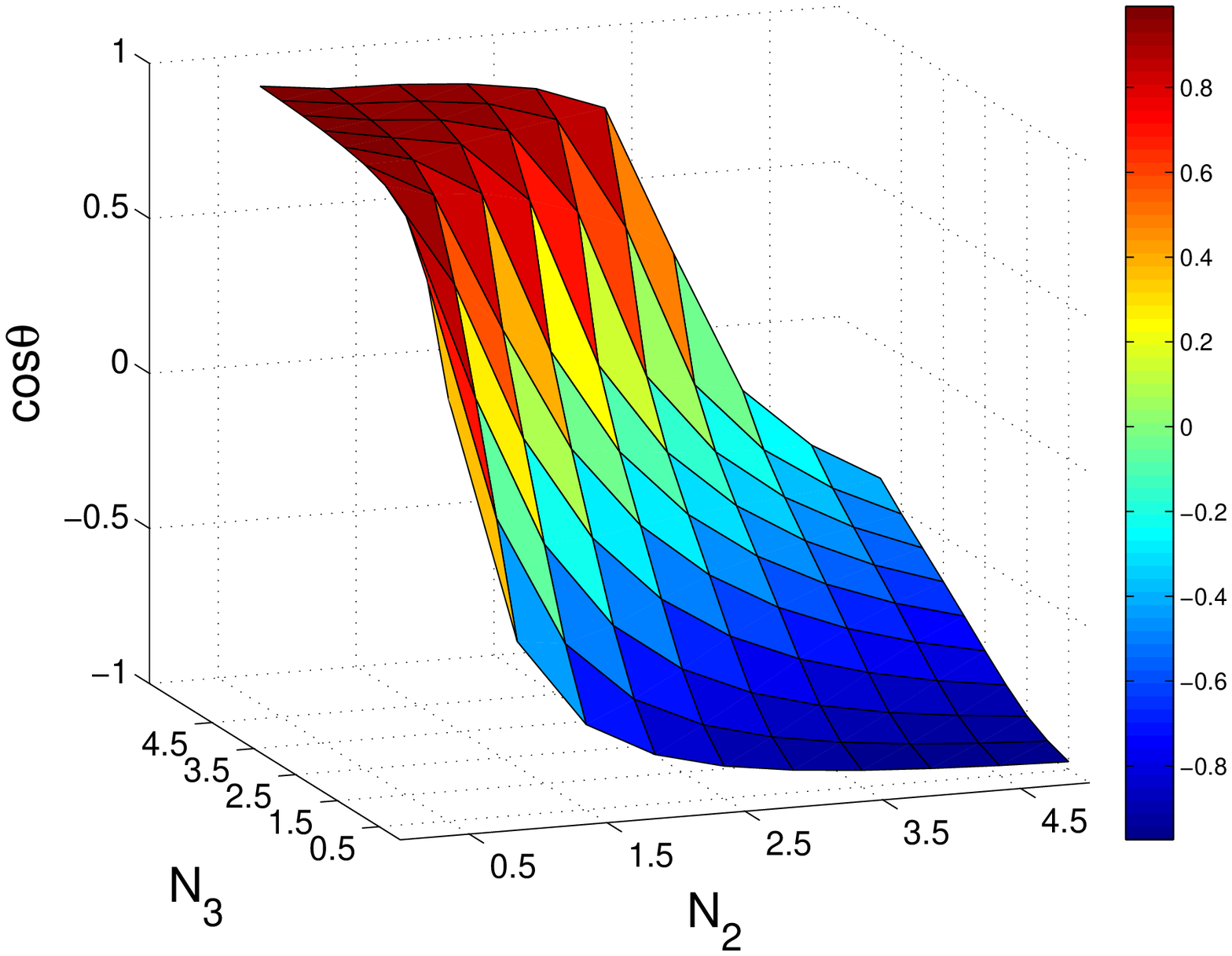}
}
\subfloat[]{\label{fig1b}%
\includegraphics[width=0.48\textwidth,keepaspectratio=true]{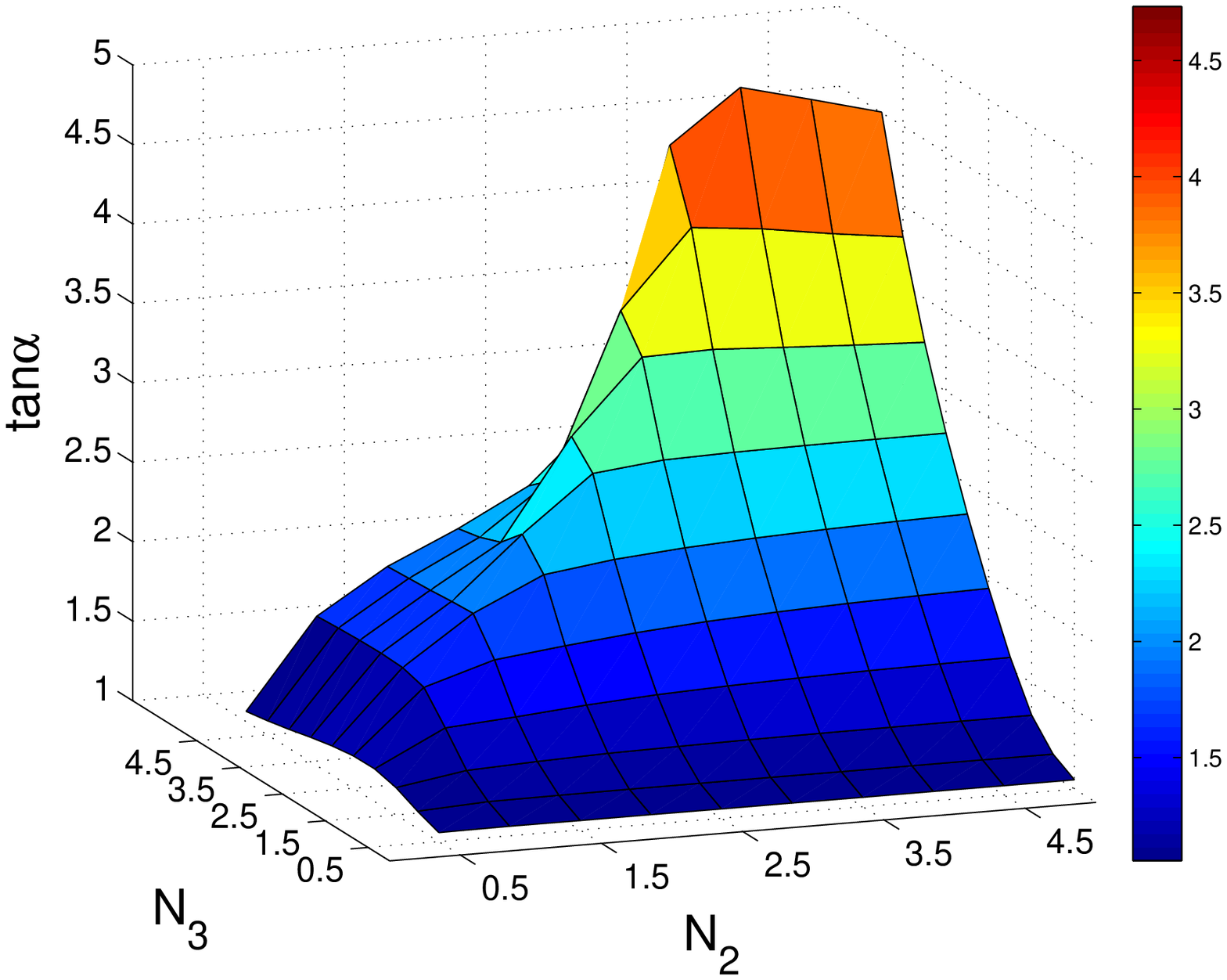}
}

\subfloat[]{\label{fig1c}%
\includegraphics[width=0.48\textwidth,keepaspectratio=true]{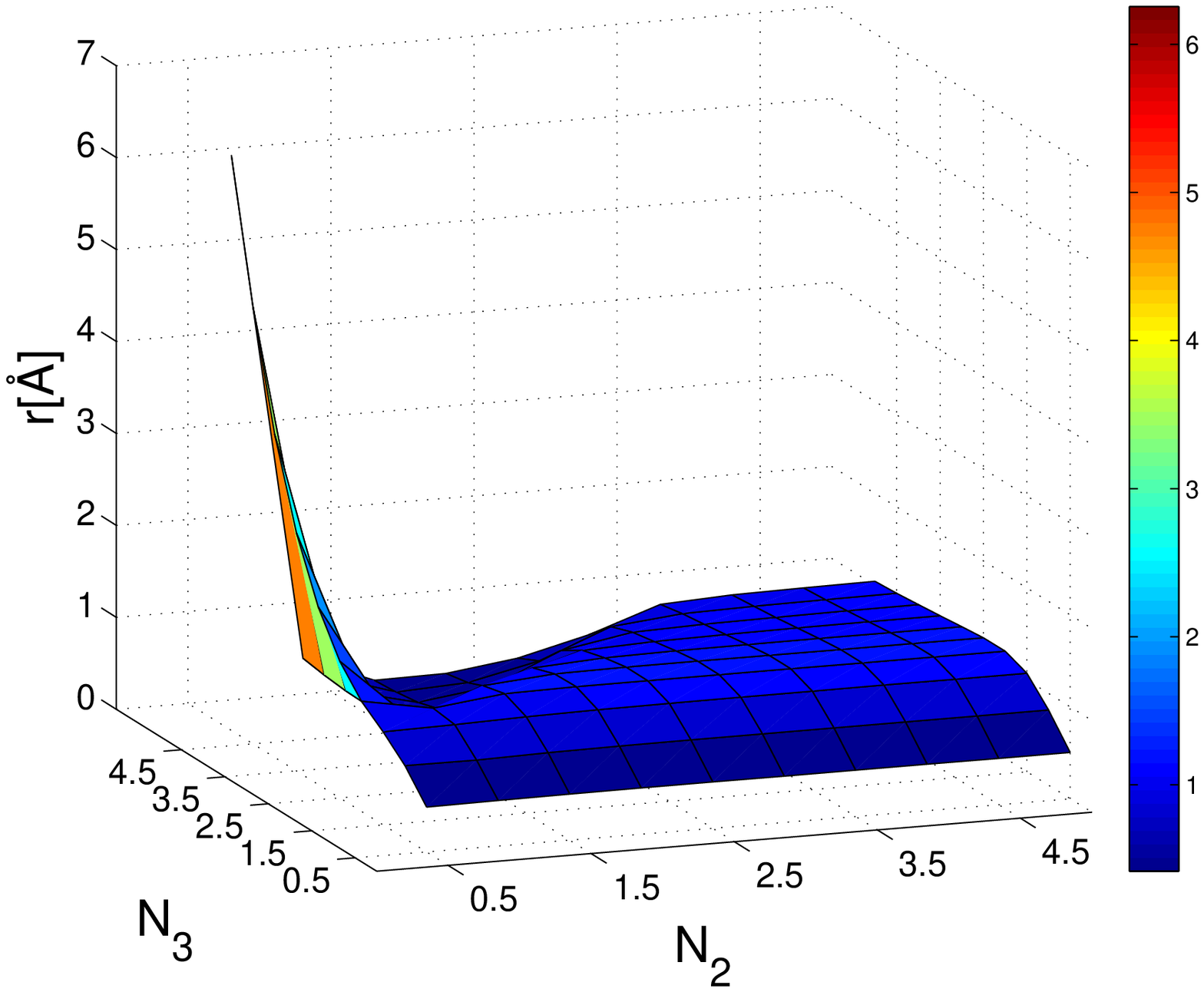}
}
\subfloat[]{\label{fig1d}%
\includegraphics[width=0.48\textwidth,keepaspectratio=true]{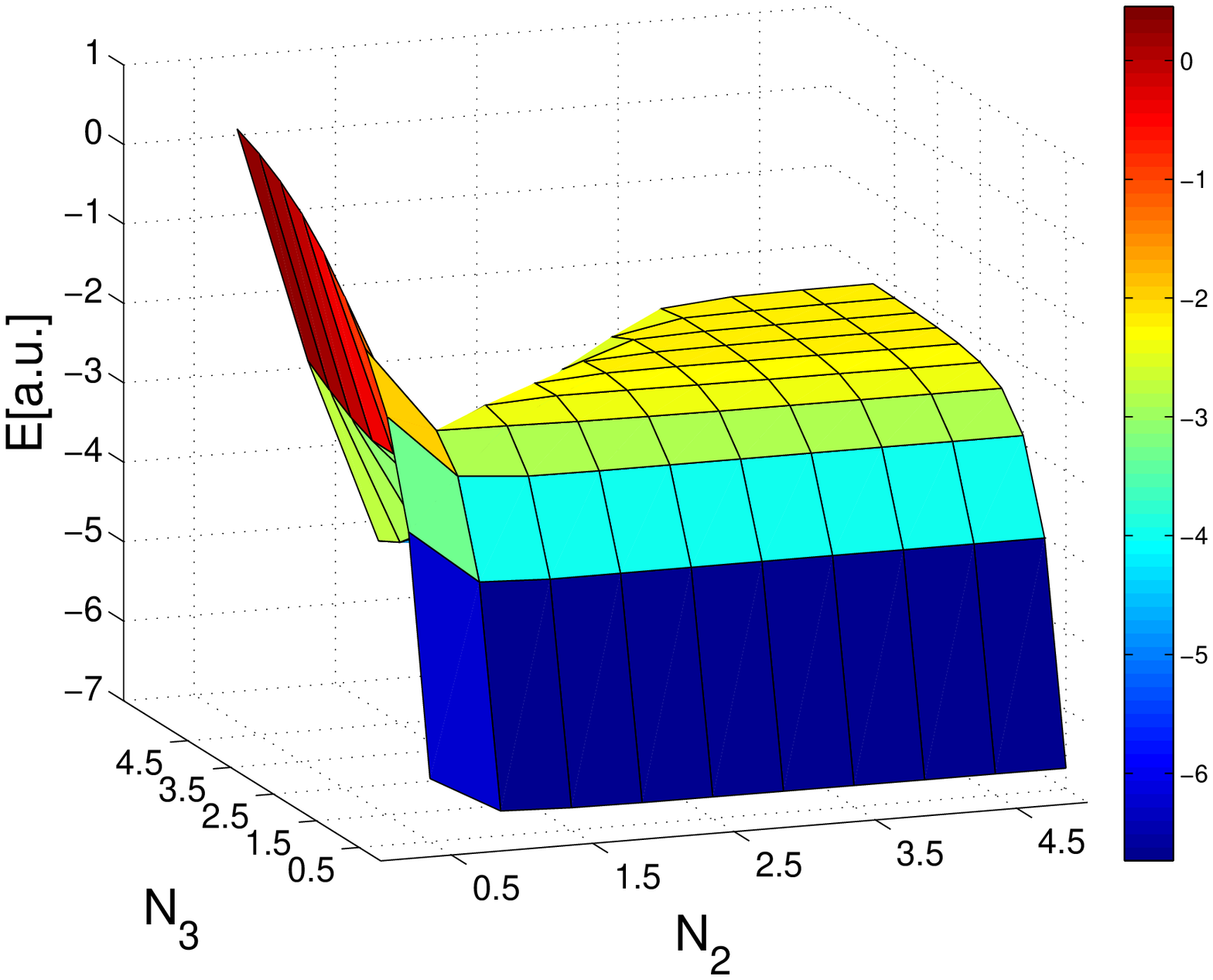}
}
\caption{For $N_1=1$ a surface plot of $\cos\theta$ (top left), $\tan\alpha$ (top right), $r$ (bottom left) and $E$ (bottom right) are shown for different values of $N_2$ and $N_3$. The data can also be found in the supporting material.}
\label{fig1}
\end{figure}

To investigate how the quantum numbers determine the energy five three dimensional surface plots are shown in Figure \ref{fig2}(a) of the energy for different integer numbers of $N_1$ (1,2,3,4 and 5) and all combinations of $N_2$ and $N_3$. Here, the surface plot of $N_1=1$ is lowest in the plot, over this plot is the surface plot of $N_1=2$ and so forth. We see that the energy is increased by a substantial amount when $N_1$ increases. Furthermore, the data suggests that the energy converges to the ionization thresholds for fixed $N_1$ and $N_2=N_3\rightarrow \infty$ and the three-particle breakup threshold for $N_1=N_2=N_3\rightarrow \infty$. In Figure \ref{fig2}(b) the corresponding surface plots of $r$ are shown. As for the energy the surface plot of $N_1=1$ is lowest in the plot, over this plot is the surface plot of $N_1=2$ and so forth. We see that $r$ increases substantially when $N_1$ increases and when $N_3$ increases in particular for high values of $N_1$. The equivalent surface plots of $\cos\theta$ and $\tan\alpha$ are not shown but the data can be found in the supporting material. The data shows that $\cos\theta$ has a little dependence on $N_1$ but is mainly determined by $N_2$ and $N_3$ while $\tan\alpha$ converges towards 1 when $N_1\rightarrow \infty$.

Next, we investigate  $\cos\theta$, $\tan\alpha$, $r$, and $E$ for $N_1$ set to 1. Four surface plots are shown in Figure \ref{fig1}. The data clearly shows that an increase in $N_2$ leads to a decrease in $\cos\theta$ while an increase in $N_3$ leads to an increase in $\cos\theta$. The results of an increase in $N_2$ ($N_3$) is thus that the motion of the two electrons becomes more parallel (antiparallel). The results suggest that an increase in $\tan\alpha$ requires an increase in both $N_2$ and $N_3$. Furthermore, $r$ is almost constant for large values of $N_2$ and $N_3$ ($r_2\approx a$) while the energy as described above converges to the first ionization threshold at $-2$ a.u.. When the system is excited by raising $N_2$ and $N_3$ simultaneously it thus follows that the angle between the electrons and the distance to the inner electron more or less are unchanged while the distance to the outer electron is increased substantially. 

Finally, the data suggests (see the supporting material for the exact values) that 1 is the lowest state for $N_1$ since the energies of $N_1=0.5$ are too low. Furthermore, it is interesting that $N_1=1$, $N_2=N_3=3/2$ gives the energy $-78.44$eV, which only is $0.57$eV from the experimental value of $-79.01$eV.

\section{Conclusion}
The object of this paper was to introduce a harmonic oscillator model to calculate the distance to the inner and outer electron, the angle between the electrons and the energy levels of helium. The method is based on a second order expansion of a two-electron Coulomb potential and the algebraic method for a harmonic oscillator. We derived a simple formula for the energy of helium in the general case and in the specific case of the Wannier ridge where $\tan\alpha=1$. A diagonalization of $\bm{C}^2$ led to three harmonic oscillators and the system can thus be described by three quantum numbers. The three harmonic oscillators are related through equation \eqref{r} which gives an equation for $r$ as a function of $\theta$ and $\alpha$. We found that the intersection point between the three surfaces uniquely determines $r$, $\cos\theta$ and $\tan\alpha$ and thus $E$ for a choice of $N_1$, $N_2$ and $N_3$ and calculated the values for 1000 different combinations of the three quantum numbers using an iterative method. 

Our analysis of the dependence of $r$, $\cos\theta$, $\tan\alpha$ and $E$ on the three quantum numbers indicates that the outer electron is excited towards the ionization threshold when both $N_2$ and $N_3$ is increased. The system becomes highly parallel or antiparallel if only one of the two quantum numbers is increased. Finally, the ground state is at $N_1=1$, $N_2=N_3=3/2$ with energy $-78.44$eV. 

\clearpage

\appendix

\section{Proof of the first relation in equation \eqref{rel}}
In this section we prove the following relation:
\begin{equation}
\sum_i\frac{1}{1+\gamma_{1,i}^2+\left(1+\tan^2\alpha\right)\gamma_{2,i}^2}=1
\end{equation}
It is easy to see that we have to prove
\begin{equation}\begin{aligned}
&\left(1+\gamma_{1,1}^2+\left(1+\tan^2\alpha\right)\gamma_{2,1}^2\right)\left(1+\gamma_{1,2}^2+\left(1+\tan^2\alpha\right)\gamma_{2,2}^2\right)\left(1+\gamma_{1,3}^2+\left(1+\tan^2\alpha\right)\gamma_{2,3}^2\right)\\
&=\left(1+\gamma_{1,1}^2+\left(1+\tan^2\alpha\right)\gamma_{2,1}^2\right)\left(1+\gamma_{1,2}^2+\left(1+\tan^2\alpha\right)\gamma_{2,2}^2\right)\\
&+\left(1+\gamma_{1,1}^2+\left(1+\tan^2\alpha\right)\gamma_{2,1}^2\right)\left(1+\gamma_{1,3}^2+\left(1+\tan^2\alpha\right)\gamma_{2,3}^2\right)\\
&+\left(1+\gamma_{1,2}^2+\left(1+\tan^2\alpha\right)\gamma_{2,2}^2\right)\left(1+\gamma_{1,3}^2+\left(1+\tan^2\alpha\right)\gamma_{2,3}^2\right).
\end{aligned}\end{equation}
This equation can be written as
\begin{equation}\begin{aligned}
&\left(\gamma_{1,1}^2+\left(1+\tan^2\alpha\right)\gamma_{2,1}^2\right)\left(\gamma_{1,2}^2+\left(\tan^2\alpha\right)\gamma_{2,2}^2\right)\left(\gamma_{1,3}^2+\left(1+\tan^2\alpha\right)\gamma_{2,3}^2\right)\\
&=\gamma_{1,1}^2\gamma_{1,2}^2\gamma_{1,3}^2+\gamma_{2,1}^2\gamma_{2,2}^2\gamma_{2,3}^2\left(1+\tan^{-2}\alpha\right)^3+\left(\gamma_{2,1}^2\gamma_{1,2}^2\gamma_{1,3}^2+\gamma_{2,2}^2\gamma_{1,1}^2\gamma_{1,3}^2+\gamma_{2,3}^2\gamma_{1,1}^2\gamma_{1,2}^2\right)\left(1+\tan^{-2}\alpha\right)\\
&+\left(\gamma_{2,1}^2\gamma_{2,2}^2\gamma_{1,3}^2+\gamma_{2,1}^2\gamma_{2,3}^2\gamma_{1,2}^2+\gamma_{2,2}^2\gamma_{2,3}^2\gamma_{1,1}^2\right)\left(1+\tan^{-2}\alpha\right)^2\\
&=-1+\sum_i\left(1+\gamma_{1,i}^2+\left(1+\tan^2\alpha\right)\gamma_{2,i}^2\right).
\end{aligned}\end{equation}
Now, using equation \eqref{ortheq} we obtain the equations:
\begin{equation}\begin{aligned}
&\gamma_{2,1}^2\gamma_{2,2}^2\gamma_{2,3}^2\left(1+\tan^{-2}\alpha\right)^3=-1-\gamma_{1,1}\gamma_{1,2}-\gamma_{1,1}\gamma_{1,3}-\gamma_{1,2}\gamma_{1,3}-\gamma_{1,1}^2\gamma_{1,2}\gamma_{1,3}-\gamma_{1,2}^2\gamma_{1,1}\gamma_{1,3}-\gamma_{1,3}^2\gamma_{1,1}\gamma_{1,2}\\
&\left(\gamma_{2,1}^2\gamma_{1,2}^2\gamma_{1,3}^2+\gamma_{2,2}^2\gamma_{1,1}^2\gamma_{1,3}^2+\gamma_{2,3}^2\gamma_{1,1}^2\gamma_{1,2}^2\right)\left(1+\tan^{-2}\alpha\right)=\gamma_{2,1}^2\gamma_{1,2}\gamma_{1,3}\left(-1-\gamma_{2,2}\gamma_{2,3} \left(1+\tan^{-2}\alpha\right)\right)\left(1+\tan^{-2}\alpha\right)\\
&+\gamma_{2,2}^2\gamma_{1,1}\gamma_{1,3}\left(-1-\gamma_{2,1}\gamma_{2,3} \left(1+\tan^{-2}\alpha\right)\right)\left(1+\tan^{-2}\alpha\right)+\gamma_{2,3}^2\gamma_{1,1}\gamma_{1,2}\left(-1-\gamma_{2,1}\gamma_{2,2} \left(1+\tan^{-2}\alpha\right)\right)\left(1+\tan^{-2}\alpha\right)\\
&\left(\gamma_{2,1}^2\gamma_{2,2}^2\gamma_{1,3}^2+\gamma_{2,1}^2\gamma_{2,3}^2\gamma_{1,2}^2+\gamma_{2,2}^2\gamma_{2,3}^2\gamma_{1,1}^2\right)\left(1+\tan^{-2}\alpha\right)^2=\gamma_{1,1}^2\gamma_{2,2}\gamma_{2,3}\left(-1-\gamma_{1,2}\gamma_{1,3}\right)\left(1+\tan^{-2}\alpha\right)\\
&+\gamma_{1,2}^2\gamma_{2,1}\gamma_{2,3}\left(-1-\gamma_{1,1}\gamma_{1,3}\right)\left(1+\tan^{-2}\alpha\right)+\gamma_{1,3}^2\gamma_{2,1}\gamma_{2,2}\left(-1-\gamma_{1,1}\gamma_{1,2}\right)\left(1+\tan^{-2}\alpha\right)\\
&\gamma_{2,1}^2\gamma_{1,2}\gamma_{1,3}+\gamma_{2,2}^2\gamma_{1,1}\gamma_{1,3}+\gamma_{2,3}^2\gamma_{1,1}\gamma_{1,2}=\gamma_{2,1}^2+\gamma_{2,2}^2+\gamma_{2,3}^2+2\left(\gamma_{2,1}^2\gamma_{2,2}\gamma_{2,2}+\gamma_{2,2}^2\gamma_{2,1}\gamma_{2,3}+\gamma_{2,3}^2\gamma_{2,1}\gamma_{2,2}\right)\left(1+\tan^{-2}\alpha\right)\\
&+3\gamma_{2,1}^2\gamma_{2,2}^2\gamma_{2,3}^2\left(1+\tan^{-2}\alpha\right)^2,
\end{aligned}\end{equation}
which when combined lead to the result.

\section{The special case of the Wannier ridge}
The eigenvalues and eigenvectors for the special case of $\alpha=\pi/4$ and $r_1=r_2=r$, the so called Wannier ridge\cite{richter1990analysis}, have to be derived separately. This important exception is treated below.

Clearly, $\lambda=0$ is still an eigenvalue with the same eigenvector. Using equation \eqref{gamma12} for $\tan\alpha=1$ we find that $\gamma_1=1$ and $\gamma_2=0$ are two solutions to the equation. When $\gamma_2=0$ we have $\gamma_1=-1$. On the other hand when $\gamma_1=1$ the cubic equation reduces to the quadratic equation:
\begin{equation}\begin{aligned}
-1+b^2-\beta b=0,
\end{aligned}\end{equation}
with
\begin{equation}\begin{aligned}
\beta&=\frac{2^4\sin^3\frac{\theta}{2}+2\cos\theta}{-\sin\theta},
\end{aligned}\end{equation}
and
\begin{equation}\begin{aligned}
b=\gamma_{\pm}.
\end{aligned}\end{equation}
The solution to the quadratic equation is
\begin{equation}\begin{aligned}
\gamma_{\pm}=\frac{\beta}{2}\left(1\pm \sqrt{1+\frac{4}{\beta^2}}\right).
\end{aligned}\end{equation}
The solution has the following properties
\begin{equation}\begin{aligned}
\gamma_+\gamma_-=-1\\
\beta=\gamma_{+}+\gamma_{-}=\gamma_+-\frac{1}{\gamma_+}=\frac{\gamma_+^2-1}{\gamma_+}=\frac{\gamma_-^2-1}{\gamma_-}\\
\gamma_+-\gamma_-=\frac{\gamma_+^2+1}{\gamma_+}=-\frac{\gamma_-^2+1}{\gamma_-}.
\end{aligned}\end{equation}
To conclude, the eigenvalues of $\bm{C}^2$ when $\tan\alpha=1$ are given by 
\begin{equation}\begin{aligned}
\lambda_1&=-2^5\sin^3\frac{\theta}{2}=-4\frac{r_{12}^3}{r^3}\\
\lambda_2&=-4\frac{r^2}{r_{12}^2}\sin\theta\frac{1-\cos\theta-\sin\theta\gamma_+}{\gamma_+}\\
\lambda_3&=-4\frac{r^2}{r_{12}^2}\sin\theta\frac{1-\cos\theta-\sin\theta\gamma_-}{\gamma_-}\\
\lambda_4&=0.
\end{aligned}\end{equation}
and the eigenvectors are given by
\begin{equation}\begin{aligned}
\bm{e}_1=\frac{1}{\sqrt{2}}\begin{pmatrix}1\\0\\-\cos\theta\\-\sin\theta \end{pmatrix},\;
\bm{e}_2=\frac{1}{\sqrt{2}}\frac{1}{\sqrt{\gamma_+^2+1}}\begin{pmatrix}1\\\gamma_+\\\cos\theta+\sin\theta\gamma_+\\\sin\theta-\cos\theta\gamma_+ \end{pmatrix},\;\bm{e}_3=\frac{1}{\sqrt{2}}\frac{1}{\sqrt{\gamma_-^2+1}}\begin{pmatrix}1\\\gamma_-\\\cos\theta+\sin\theta\gamma_-\\\sin\theta-\cos\theta\gamma_- \end{pmatrix},\;
\bm{e}_4=\frac{1}{\sqrt{2}}\begin{pmatrix}0\\1\\-\sin\theta\\\cos\theta \end{pmatrix}.
\end{aligned}\end{equation}
The reader can readily confirm that the eigenvectors are orthogonal. 

Next, we calculate $c_i=\bm{e}_i^T\bm{c}_i$:
\begin{equation}\begin{aligned}
c_1&=-\sqrt{2}2^4\sin^3\frac{\theta}{2}r=-\frac{2\sqrt{2}r_{12}^3}{r^2}\\
c_2&=-\sqrt{2}r\frac{1-\cos\theta-\sin\theta\gamma_+}{\sqrt{\gamma_+^2+1}}\\
c_3&=-\sqrt{2}r\frac{1-\cos\theta-\sin\theta\gamma_-}{\sqrt{\gamma_-^2+1}}\\
c_4&=0.
\end{aligned}\end{equation}

The formulas for the eigenvalues and the eigenvalues and the $c_i$'s are so simple that we readily can derive a formula for $r$ and for the different energy terms. Using equation \eqref{r} for $c_1$ and $\lambda_1$ we obtain 
\begin{equation}\begin{aligned}
r_{12}^3=4\frac{\left(-2\sqrt{2}\frac{r_{12}^3}{r^2}\right)^4}{N_1^2\left(-4\frac{r_{12}^3}{r^3}\right)^3a}.
\end{aligned}\end{equation}
Hence, we find that $r$ is given by
\begin{equation}\begin{aligned}
r=-N_1^2a=n_1^2a,
\end{aligned}\end{equation}
where $N_1=in_1$. 
To use equation \eqref{H}, we calculate
\begin{equation}\begin{aligned}
\frac{c_1^2}{\lambda_1}&=-2\frac{r_{12}^3}{r}\\
\frac{c_2^2}{\lambda_2}&=-\frac{r_{12}^2}{2}\frac{1}{\sin\theta}\frac{\gamma_+}{\gamma_+^2+1}\left(1-\cos\theta-\sin\theta\gamma_+\right)\\
\frac{c_3^2}{\lambda_3}&=-\frac{r_{12}^2}{2}\frac{1}{\sin\theta}\frac{\gamma_-}{\gamma_-^2+1}\left(1-\cos\theta-\sin\theta\gamma_-\right)
\end{aligned}\end{equation}
Finally using $\frac{\gamma_-}{\gamma_-^2+1}=-\frac{1}{\gamma_+-\gamma_-}$ and $\frac{\gamma_+}{\gamma_+^2+1}=\frac{1}{\gamma_+-\gamma_-}$ we obtain the formula for the total energy of the helium atom at the Wannier ridge:
\begin{equation}\begin{aligned}
E&=2E_0\frac{2a}{r_{12}^3}\sum_i\frac{c_i^2}{\lambda_i}\\
&=2E_0\frac{2a}{r}\left(-2+\frac{r}{2r_{12}}\right).
\end{aligned}\end{equation}

\bibliographystyle{apsrev4-1} 
\bibliography{main}

\newpage

\section{Supporting material}
Numerical data for $r$, $\cos\theta$, $\tan\alpha$, $E$ and the residuals. The value of $N_1$ is in the first row and first column. The value of $N_2$ is in first row from the second to the eleventh column. The value of $N_3$ is in the first column from the second to the eleventh row. The data is tabulated from the second to the eleventh row and from the second to the eleventh column. The data was calculated using the iterative intersection method described in main document. 

\clearpage
\subsection{The distance to the inner electron - $r$ in Angstrom}
\clearpage
\begin{table}
\begin{tabular}{|c|c|c|c|c|c|c|c|c|c|c|}
\hline
0.5&0.5&1&1.5&2&2.5&3&3.5&4&4.5&5\\\hline
0.5&0.197&0.20277&0.20418&0.20468&0.20492&0.20505&0.20512&0.20517&0.20521&0.20523\\\hline
1&0.26474&0.29101&0.29164&0.29152&0.29139&0.29131&0.29125&0.29122&0.29119&0.29117\\\hline
1.5&0.16824&0.26868&0.28316&0.28448&0.28467&0.28467&0.28464&0.28461&0.28458&0.28456\\\hline
2&0.16388&0.14321&0.25997&0.26861&0.27117&0.27235&0.27298&0.27337&0.27363&0.2738\\\hline
2.5&0.18006&0.10637&0.18006&0.25609&0.2608&0.26304&0.26436&0.26521&0.2658&0.26621\\\hline
3&0.20569&0.093861&0.13046&0.20526&0.25403&0.25669&0.25834&0.25948&0.26029&0.26089\\\hline
3.5&0.23935&0.087936&0.11058&0.15766&0.21666&0.25278&0.25438&0.25555&0.25643&0.25712\\\hline
4&0.28259&0.084672&0.10009&0.13427&0.17753&0.24858&0.25209&0.253&0.25381&0.25449\\\hline
4.5&0.33989&0.082701&0.093741&0.12068&0.15471&0.19034&0.22247&0.24153&0.25276&0.25293\\\hline
5&0.42506&0.08143&0.089566&0.11197&0.14005&0.17007&0.19848&0.22276&0.24213&0.24262\\\hline
\end{tabular}
\end{table}
\begin{table}\begin{tabular}{|c|c|c|c|c|c|c|c|c|c|c|}
\hline
1&0.5&1&1.5&2&2.5&3&3.5&4&4.5&5\\\hline
0.5&0.48825&0.49568&0.50073&0.50266&0.50358&0.50408&0.50438&0.50458&0.50471&0.50481\\\hline
1&0.7923&0.788&0.8037&0.81107&0.8147&0.81671&0.81793&0.81873&0.81928&0.81967\\\hline
1.5&1.0127&1.0259&1.0481&1.0595&1.0652&1.0683&1.0702&1.0714&1.0722&1.0728\\\hline
2&1.1968&1.059&1.1521&1.164&1.1663&1.1666&1.1663&1.1661&1.1658&1.1656\\\hline
2.5&1.4868&0.79027&1.0932&1.1566&1.1658&1.1668&1.1662&1.1653&1.1646&1.1639\\\hline
3&1.9304&0.67297&0.75021&1.0747&1.1212&1.1327&1.1365&1.1379&1.1385&1.1387\\\hline
3.5&2.5843&0.64693&0.54307&0.82873&1.0539&1.0876&1.0997&1.1055&1.1087&1.1106\\\hline
4&3.5147&0.65551&0.47032&0.57284&0.90557&1.0399&1.0636&1.0744&1.0807&1.0847\\\hline
4.5&4.7524&0.68204&0.43573&0.47563&0.64594&0.95537&1.0307&1.0472&1.0562&1.0621\\\hline
5&6.2552&0.72026&0.41726&0.42547&0.52912&0.72024&0.98624&1.0244&1.036&1.0432\\\hline
\end{tabular}
\end{table}
\begin{table}\begin{tabular}{|c|c|c|c|c|c|c|c|c|c|c|}
\hline
1.5&0.5&1&1.5&2&2.5&3&3.5&4&4.5&5\\\hline
0.5&0.91144&0.89901&0.90584&0.91073&0.91313&0.91445&0.91526&0.91578&0.91614&0.9164\\\hline
1&1.3913&1.3142&1.3342&1.3483&1.3558&1.36&1.3626&1.3643&1.3655&1.3663\\\hline
1.5&1.9248&1.7549&1.773&1.7992&1.8153&1.8249&1.8309&1.8349&1.8376&1.8396\\\hline
2&2.4821&2.1271&2.1592&2.192&2.2153&2.2299&2.2392&2.2454&2.2497&2.2528\\\hline
2.5&3.1408&2.2522&2.4099&2.4608&2.4849&2.4985&2.5065&2.5116&2.5149&2.5172\\\hline
3&3.9755&2.2162&2.3827&2.5657&2.6065&2.6191&2.6233&2.6246&2.6247&2.6245\\\hline
3.5&5.0122&2.2451&1.9694&2.4563&2.5954&2.6287&2.6375&2.639&2.6382&2.6368\\\hline
4&6.2466&2.3653&1.65&1.971&2.4506&2.5586&2.5903&2.601&2.6046&2.6054\\\hline
4.5&7.6598&2.5636&1.5142&1.4703&2.0606&2.4181&2.5038&2.5348&2.5485&2.5551\\\hline
5&&2.835&1.4633&1.2551&1.4896&2.1396&2.3853&2.4529&2.4821&2.4973\\\hline
\end{tabular}
\end{table}
\begin{table}\begin{tabular}{|c|c|c|c|c|c|c|c|c|c|c|}
\hline
2&0.5&1&1.5&2&2.5&3&3.5&4&4.5&5\\\hline
0.5&1.4807&1.4114&1.4244&1.4339&1.439&1.4418&1.4435&1.4447&1.4455&1.446\\\hline
1&2.1216&1.953&1.9619&1.9827&1.9954&2.0029&2.0076&2.0107&2.0128&2.0143\\\hline
1.5&2.8685&2.5565&2.5381&2.5693&2.5937&2.6092&2.6192&2.6258&2.6305&2.6338\\\hline
2&3.7031&3.1692&3.1193&3.152&3.1887&3.2148&3.2323&3.2443&3.2527&3.2588\\\hline
2.5&4.6352&3.6838&3.6487&3.6827&3.7263&3.7611&3.7858&3.803&3.8153&3.8242\\\hline
3&5.7013&4.0509&4.0112&4.1035&4.1547&4.1924&4.2193&4.238&4.2512&4.2608\\\hline
3.5&6.9224&4.3843&4.0464&4.3326&4.4329&4.4772&4.5024&4.5179&4.5278&4.5345\\\hline
4&8.2999&4.7873&3.8361&4.2359&4.5215&4.6086&4.6418&4.6561&4.6625&4.6652\\\hline
4.5&9.8276&5.3013&3.6672&3.7095&4.3511&4.5827&4.6568&4.6839&4.6936&4.6963\\\hline
5&11.4984&5.9471&3.6151&3.1611&3.7639&4.3727&4.5601&4.6266&4.6529&4.6633\\\hline
\end{tabular}
\end{table}
\begin{table}\begin{tabular}{|c|c|c|c|c|c|c|c|c|c|c|}
\hline
2.5&0.5&1&1.5&2&2.5&3&3.5&4&4.5&5\\\hline
0.5&2.1863&2.0183&2.0538&2.0722&2.0811&2.0864&2.0897&2.0918&2.0933&2.0944\\\hline
1&2.9726&2.7268&2.7018&2.7257&2.7447&2.7566&2.7643&2.7694&2.773&2.7755\\\hline
1.5&3.8802&3.4876&3.4045&3.4294&3.4609&3.4832&3.498&3.5081&3.5152&3.5203\\\hline
2&4.9003&4.3029&4.1466&4.1575&4.1993&4.2345&4.2595&4.277&4.2894&4.2985\\\hline
2.5&6.0292&5.1152&4.8901&4.8799&4.925&4.9721&5.0087&5.0354&5.0548&5.0692\\\hline
3&7.2752&5.8734&5.5557&5.5544&5.5989&5.6531&5.6991&5.7343&5.7606&5.7803\\\hline
3.5&8.6519&6.5991&6.0218&6.1102&6.1763&6.2352&6.2858&6.3254&6.3554&6.3781\\\hline
4&10.1662&7.3719&6.2432&6.4298&6.6002&6.6815&6.736&6.7756&6.8046&6.826\\\hline
4.5&11.8187&8.2589&6.3556&6.3784&6.7888&6.9563&7.032&7.0746&7.1012&7.1186\\\hline
5&13.6072&9.3022&6.4989&6.0194&6.6186&7.0142&7.1626&7.2272&7.2588&7.2752\\\hline
\end{tabular}
\end{table}
\begin{table}\begin{tabular}{|c|c|c|c|c|c|c|c|c|c|c|}
\hline
3&0.5&1&1.5&2&2.5&3&3.5&4&4.5&5\\\hline
0.5&3.0162&2.8367&2.8404&2.7572&2.8392&2.8479&2.8534&2.8572&2.8597&2.8616\\\hline
1&3.9374&3.6458&3.5616&3.596&3.6059&3.6234&3.635&3.6429&3.6485&3.6525\\\hline
1.5&4.9841&4.5636&4.3942&4.3971&4.4318&4.4611&4.4818&4.4963&4.5065&4.514\\\hline
2&6.1515&5.5653&5.2888&5.2568&5.2948&5.3369&5.3696&5.3932&5.4104&5.4231\\\hline
2.5&7.4366&6.6214&6.2191&6.141&6.1715&6.2239&6.2698&6.305&6.3312&6.3509\\\hline
3&8.8392&7.6992&7.1307&7.0197&7.0355&7.092&7.1495&7.1968&7.2333&7.2613\\\hline
3.5&10.3637&8.7897&7.9425&7.839&7.8523&7.9084&7.9731&8.0301&8.0761&8.1121\\\hline
4&12.0158&9.9284&8.593&8.5082&8.5718&8.6366&8.7053&8.7678&8.8197&8.8611\\\hline
4.5&13.7984&11.171&9.1145&8.9111&9.1178&9.2329&9.3134&9.3794&9.433&9.4756\\\hline
5&15.713&12.5631&9.6048&9.0087&9.3793&9.6398&9.7654&9.8432&9.8985&9.9395\\\hline
\end{tabular}
\end{table}
\begin{table}\begin{tabular}{|c|c|c|c|c|c|c|c|c|c|c|}
\hline
3.5&0.5&1&1.5&2&2.5&3&3.5&4&4.5&5\\\hline
0.5&3.9665&3.6993&3.6817&3.7561&3.8052&3.8043&3.7344&3.7403&3.7444&3.7474\\\hline
1&5.0179&4.7125&4.5476&4.5373&4.5802&4.6036&4.6202&4.6317&4.6399&4.6459\\\hline
1.5&6.1958&5.7852&5.5194&5.4819&5.5128&5.5479&5.5751&5.5948&5.609&5.6195\\\hline
2&7.4972&6.9618&6.5658&6.4708&6.4933&6.5387&6.5787&6.6091&6.6318&6.6487\\\hline
2.5&8.9205&8.2261&7.6721&7.5064&7.5065&7.5565&7.6092&7.6526&7.686&7.7114\\\hline
3&10.4644&9.5596&8.801&8.5658&8.5338&8.5801&8.6428&8.6994&8.7451&8.7809\\\hline
3.5&12.13&10.9496&9.9001&9.6081&9.5496&9.5856&9.653&9.7208&9.7788&9.8256\\\hline
4&13.9184&12.4043&10.9203&10.5658&10.514&10.5439&10.6116&10.6866&10.7545&10.8113\\\hline
4.5&15.8337&13.9543&11.8517&11.3503&11.3641&11.4163&11.4877&11.5666&11.6405&11.7043\\\hline
5&17.8765&15.6346&12.7397&11.9013&12.0074&12.1483&12.2456&12.3318&12.4093&12.4761\\\hline
\end{tabular}
\end{table}
\begin{table}\begin{tabular}{|c|c|c|c|c|c|c|c|c|c|c|}
\hline
4&0.5&1&1.5&2&2.5&3&3.5&4&4.5&5\\\hline
0.5&5.0377&4.8204&4.7053&4.5539&4.6494&4.7124&4.7342&4.741&4.7474&4.7517\\\hline
1&6.2172&5.9228&5.6874&5.6455&5.6775&5.6977&5.7198&5.7357&5.7473&5.7558\\\hline
1.5&7.5231&7.1465&6.7877&6.6917&6.7079&6.7459&6.7795&6.8051&6.8242&6.8384\\\hline
2&8.9537&8.4862&7.987&7.812&7.805&7.8476&7.8934&7.9309&7.9597&7.9817\\\hline
2.5&10.5079&9.9323&9.2645&8.9962&8.9497&8.9881&9.0441&9.0948&9.1356&9.1674\\\hline
3&12.1852&11.4739&10.5947&10.226&10.1288&10.1523&10.2138&10.277&10.3311&10.3748\\\hline
3.5&13.9852&13.1012&11.9446&11.4695&11.3226&11.3229&11.3829&11.4559&11.5231&11.5795\\\hline
4&15.9018&14.8125&13.2796&12.6769&12.4992&12.4771&12.5299&12.608&12.6861&12.7548\\\hline
4.5&17.9537&16.6189&14.5795&13.7839&13.6082&13.5832&13.6295&13.7086&13.7938&13.8725\\\hline
5&20.1266&18.5408&15.8569&14.7351&14.5762&14.5948&14.65&14.7308&14.82&14.9051\\\hline
\end{tabular}
\end{table}
\begin{table}\begin{tabular}{|c|c|c|c|c|c|c|c|c|c|c|}
\hline
4.5&0.5&1&1.5&2&2.5&3&3.5&4&4.5&5\\\hline
0.5&6.2311&6.0361&5.768&5.7412&5.7827&5.8762&5.8497&5.8582&5.8774&5.8897\\\hline
1&7.5373&7.2694&6.9525&6.9004&6.9032&6.987&6.9338&6.9548&6.9704&6.9821\\\hline
1.5&8.9705&8.6387&8.2029&8.0333&8.0233&8.0911&8.0963&8.1281&8.1526&8.1712\\\hline
2&10.5284&10.1308&9.556&9.2888&9.2374&9.2694&9.3174&9.3615&9.3969&9.4245\\\hline
2.5&12.2103&11.7389&11.0013&10.6221&10.5134&10.5292&10.5832&10.6396&10.6877&10.7264\\\hline
3&14.0169&13.4565&12.522&12.0178&11.8401&11.8277&11.8801&11.9467&12.0081&12.0595\\\hline
3.5&15.9462&15.2769&14.0973&13.4508&13.2014&13.1513&13.1936&13.266&13.3398&13.4047\\\hline
4&&17.1962&15.7039&14.8845&14.5714&14.4822&14.5072&14.5798&14.6631&14.7408\\\hline
4.5&&19.2164&17.3232&16.2732&15.9098&15.7942&15.8007&15.8684&15.957&16.0453\\\hline
5&&21.3465&18.9527&17.5728&17.1593&17.0488&17.0475&17.1092&17.1994&17.2949\\\hline
\end{tabular}
\end{table}
\begin{table}\begin{tabular}{|c|c|c|c|c|c|c|c|c|c|c|}
\hline
5&0.5&1&1.5&2&2.5&3&3.5&4&4.5&5\\\hline
0.5&7.5478&7.3746&7.185&6.7936&6.9716&6.9912&7.0834&7.0723&7.0996&7.1142\\\hline
1&8.9803&8.7451&8.3754&8.0733&8.1195&8.2153&8.2626&8.2888&8.3091&8.3245\\\hline
1.5&10.5392&10.2545&9.7666&9.498&9.5475&9.4872&9.5298&9.5642&9.5945&9.6182\\\hline
2&12.2241&11.8903&11.2722&10.9073&10.7972&10.807&10.8537&10.9028&10.9449&10.9787\\\hline
2.5&14.0321&13.6472&12.8819&12.391&12.2062&12.1873&12.2326&12.2916&12.3463&12.392\\\hline
3&&15.5206&14.5838&13.9505&13.6796&13.6182&13.6526&13.7178&13.7847&13.8435\\\hline
3.5&&17.5066&16.3646&15.5662&15.2036&15.0894&15.1031&15.1684&15.2455&15.3176\\\hline
4&&19.6012&18.2089&17.2115&16.7565&16.5865&16.5714&16.6298&16.7131&16.7973\\\hline
4.5&&21.8039&20.1022&18.854&18.3059&18.0876&18.041&18.0864&18.1709&18.2644\\\hline
5&&24.117&22.0373&20.4606&19.8076&19.5605&19.4896&19.5197&19.6009&19.7\\\hline
\end{tabular}
\end{table}
\clearpage
\subsection{The angle between the electrons - $\cos\theta$}
\clearpage
\begin{table}\begin{tabular}{|c|c|c|c|c|c|c|c|c|c|c|}
\hline
0.5&0.5&1&1.5&2&2.5&3&3.5&4&4.5&5\\\hline
0.5&-0.082998&-0.70383&-0.86231&-0.92135&-0.94931&-0.96466&-0.97398&-0.98005&-0.98422&-0.98721\\\hline
1&0.58678&-0.26771&-0.62124&-0.7752&-0.85253&-0.89621&-0.92313&-0.94084&-0.95309&-0.96191\\\hline
1.5&0.90507&0.25099&-0.31316&-0.57422&-0.71378&-0.79564&-0.84727&-0.88174&-0.90583&-0.92329\\\hline
2&0.94921&0.87727&0.049786&-0.37733&-0.57815&-0.69533&-0.77013&-0.82072&-0.85646&-0.88259\\\hline
2.5&0.96571&0.94411&0.83001&-0.11693&-0.43767&-0.59679&-0.69487&-0.76085&-0.8076&-0.84196\\\hline
3&0.97431&0.9652&0.92722&0.80271&-0.24125&-0.48586&-0.61628&-0.70004&-0.75842&-0.80113\\\hline
3.5&0.97954&0.97564&0.95508&0.9128&0.80579&-0.32931&-0.52279&-0.63305&-0.70627&-0.75865\\\hline
4&0.98299&0.9818&0.96852&0.94526&0.90586&0.35851&-0.40233&-0.55106&-0.64667&-0.71199\\\hline
4.5&0.98538&0.98582&0.9764&0.96117&0.93865&0.90564&0.85016&0.76034&-0.60718&-0.6669\\\hline
5&0.98704&0.98861&0.98151&0.97058&0.95553&0.93562&0.90958&0.87149&0.78525&0.80564\\\hline
\end{tabular}
\end{table}
\begin{table}\begin{tabular}{|c|c|c|c|c|c|c|c|c|c|c|}
\hline
1&0.5&1&1.5&2&2.5&3&3.5&4&4.5&5\\\hline
0.5&0.36351&-0.43969&-0.72568&-0.84054&-0.8964&-0.92747&-0.94646&-0.95888&-0.96744&-0.97358\\\hline
1&0.69917&-0.082998&-0.50667&-0.70383&-0.80481&-0.86231&-0.89789&-0.92135&-0.9376&-0.94931\\\hline
1.5&0.86433&0.26324&-0.22725&-0.50837&-0.667&-0.76161&-0.82166&-0.86185&-0.88998&-0.91037\\\hline
2&0.92295&0.58678&0.071609&-0.26771&-0.48326&-0.62124&-0.71251&-0.7752&-0.8198&-0.85253\\\hline
2.5&0.9489&0.83294&0.38182&-0.022741&-0.28759&-0.46596&-0.5885&-0.67489&-0.73749&-0.78402\\\hline
3&0.96401&0.90507&0.77714&0.25099&-0.091993&-0.31316&-0.46553&-0.57422&-0.65392&-0.71378\\\hline
3.5&0.97436&0.9337&0.8954&0.69564&0.14577&-0.1548&-0.34447&-0.47641&-0.5727&-0.64517\\\hline
4&0.98202&0.94921&0.93142&0.87727&0.59763&0.049786&-0.2131&-0.37733&-0.49262&-0.57815\\\hline
4.5&0.98767&0.95898&0.94952&0.92268&0.8534&0.48828&-0.038512&-0.26582&-0.40887&-0.51045\\\hline
5&0.99158&0.96571&0.96054&0.94411&0.91041&0.83001&0.34179&-0.11693&-0.31225&-0.43767\\\hline
\end{tabular}
\end{table}
\begin{table}\begin{tabular}{|c|c|c|c|c|c|c|c|c|c|c|}
\hline
1.5&0.5&1&1.5&2&2.5&3&3.5&4&4.5&5\\\hline
0.5&0.6531&-0.14382&-0.54953&-0.73123&-0.82332&-0.87553&-0.90776&-0.92898&-0.94367&-0.95425\\\hline
1&0.83032&0.15769&-0.32365&-0.58062&-0.71954&-0.80062&-0.85146&-0.88525&-0.90878&-0.92579\\\hline
1.5&0.91134&0.41103&-0.082998&-0.40034&-0.58883&-0.70383&-0.77764&-0.82738&-0.86231&-0.88772\\\hline
2&0.94812&0.61809&0.15188&-0.19679&-0.42909&-0.58041&-0.68117&-0.75057&-0.80001&-0.83632\\\hline
2.5&0.96694&0.77384&0.37119&0.014617&-0.24681&-0.43086&-0.55982&-0.6516&-0.71841&-0.7682\\\hline
3&0.97807&0.8582&0.58678&0.22154&-0.058529&-0.26771&-0.42202&-0.53603&-0.62124&-0.68596\\\hline
3.5&0.98517&0.89933&0.77857&0.43852&0.12821&-0.10425&-0.28075&-0.41503&-0.51778&-0.59722\\\hline
4&0.98982&0.92239&0.8667&0.70217&0.33318&0.060852&-0.14101&-0.29526&-0.41464&-0.50802\\\hline
4.5&0.99287&0.93706&0.90507&0.84906&0.61588&0.25099&0.0044592&-0.17565&-0.31316&-0.42055\\\hline
5&&0.94725&0.92627&0.89934&0.82659&0.52945&0.17947&-0.046979&-0.20977&-0.33366\\\hline
\end{tabular}
\end{table}
\begin{table}\begin{tabular}{|c|c|c|c|c|c|c|c|c|c|c|}
\hline
2&0.5&1&1.5&2&2.5&3&3.5&4&4.5&5\\\hline
0.5&0.81882&0.1203&-0.35437&-0.60151&-0.73396&-0.81102&-0.85926&-0.8913&-0.9136&-0.92972\\\hline
1&0.90659&0.36351&-0.13239&-0.43969&-0.61805&-0.72568&-0.79439&-0.84054&-0.8729&-0.8964\\\hline
1.5&0.94766&0.5537&0.079939&-0.26405&-0.48457&-0.62444&-0.71612&-0.77866&-0.82294&-0.85532\\\hline
2&0.96837&0.69917&0.27188&-0.082998&-0.3357&-0.50667&-0.62285&-0.70383&-0.76194&-0.80481\\\hline
2.5&0.97969&0.80156&0.44227&0.094581&-0.17605&-0.37295&-0.51323&-0.61393&-0.68758&-0.74263\\\hline
3&0.9864&0.86433&0.59528&0.26324&-0.012872&-0.22725&-0.38833&-0.50837&-0.59846&-0.667\\\hline
3.5&0.99061&0.90061&0.72659&0.4246&0.1473&-0.077052&-0.25371&-0.39069&-0.49661&-0.579\\\hline
4&0.99336&0.92295&0.81608&0.58678&0.30468&0.071609&-0.11684&-0.26771&-0.38774&-0.48326\\\hline
4.5&0.99519&0.93799&0.86651&0.73993&0.47078&0.22019&0.018649&-0.14489&-0.27755&-0.38508\\\hline
5&0.99644&0.9489&0.89597&0.83294&0.66205&0.38182&0.15661&-0.022741&-0.16846&-0.28759\\\hline
\end{tabular}
\end{table}
\begin{table}\begin{tabular}{|c|c|c|c|c|c|c|c|c|c|c|}
\hline
2.5&0.5&1&1.5&2&2.5&3&3.5&4&4.5&5\\\hline
0.5&0.90456&0.32805&-0.15897&-0.45872&-0.63167&-0.73566&-0.80194&-0.84643&-0.8776&-0.90025\\\hline
1&0.94729&0.52678&0.047679&-0.2923&-0.50647&-0.64115&-0.72905&-0.78888&-0.83119&-0.86209\\\hline
1.5&0.96866&0.66921&0.23113&-0.12372&-0.37067&-0.53494&-0.64549&-0.7221&-0.77688&-0.81721\\\hline
2&0.98023&0.77236&0.39015&0.03975&-0.22798&-0.41812&-0.55114&-0.64548&-0.71392&-0.7648\\\hline
2.5&0.9869&0.84228&0.52712&0.19285&-0.082998&-0.29263&-0.44624&-0.55843&-0.64138&-0.70383\\\hline
3&0.99098&0.88735&0.64456&0.33389&0.05969&-0.16146&-0.33186&-0.46079&-0.55845&-0.63319\\\hline
3.5&0.9936&0.91602&0.74071&0.46437&0.19692&-0.028592&-0.21056&-0.35366&-0.46521&-0.55236\\\hline
4&0.99535&0.9349&0.8107&0.58667&0.3283&0.10238&-0.086283&-0.24003&-0.36357&-0.4624\\\hline
4.5&0.99654&0.9481&0.85638&0.6979&0.45669&0.23014&0.0373&-0.12401&-0.25718&-0.36629\\\hline
5&0.99738&0.9579&0.88588&0.78416&0.58678&0.35736&0.15904&-0.0087104&-0.1498&-0.26771\\\hline
\end{tabular}
\end{table}
\begin{table}\begin{tabular}{|c|c|c|c|c|c|c|c|c|c|c|}
\hline
3&0.5&1&1.5&2&2.5&3&3.5&4&4.5&5\\\hline
0.5&0.94707&0.5096&0.033159&-0.32338&-0.52005&-0.65133&-0.73684&-0.795&-0.8361&-0.86612\\\hline
1&0.96873&0.6531&0.20757&-0.14382&-0.38925&-0.54953&-0.65695&-0.73123&-0.78429&-0.82332\\\hline
1.5&0.98037&0.75777&0.36351&0.011252&-0.25383&-0.43969&-0.5687&-0.65977&-0.72568&-0.77459\\\hline
2&0.98707&0.83032&0.49557&0.15769&-0.11768&-0.32365&-0.47269&-0.58062&-0.66&-0.71954\\\hline
2.5&0.99114&0.87895&0.60695&0.29086&0.015331&-0.20384&-0.36995&-0.49399&-0.58704&-0.65777\\\hline
3&0.99373&0.91134&0.69917&0.41103&0.14228&-0.082998&-0.26197&-0.40034&-0.50667&-0.58883\\\hline
3.5&0.99543&0.93311&0.77221&0.51975&0.26178&0.036231&-0.15074&-0.3007&-0.41914&-0.5125\\\hline
4&0.9966&0.94812&0.82638&0.61809&0.3741&0.15188&-0.038653&-0.19679&-0.32539&-0.42909\\\hline
4.5&0.99742&0.95889&0.86433&0.70442&0.48069&0.26324&0.072133&-0.090931&-0.22725&-0.33981\\\hline
5&0.99801&0.96694&0.89056&0.77384&0.58292&0.37119&0.18043&0.014617&-0.12714&-0.24681\\\hline
\end{tabular}
\end{table}
\begin{table}\begin{tabular}{|c|c|c|c|c|c|c|c|c|c|c|}
\hline
3.5&0.5&1&1.5&2&2.5&3&3.5&4&4.5&5\\\hline
0.5&0.96875&0.63407&0.18951&-0.15363&-0.39091&-0.55202&-0.66516&-0.73773&-0.78953&-0.82763\\\hline
1&0.98043&0.74863&0.34497&-0.009685&-0.27039&-0.45317&-0.57949&-0.66847&-0.73279&-0.78047\\\hline
1.5&0.98713&0.82349&0.4768&0.13618&-0.13838&-0.34162&-0.48765&-0.59297&-0.67023&-0.7281\\\hline
2&0.9912&0.87418&0.58632&0.26606&-0.01019&-0.2275&-0.39059&-0.51154&-0.60188&-0.67036\\\hline
2.5&0.99378&0.90837&0.67681&0.38208&0.11151&-0.11322&-0.28969&-0.42481&-0.5279&-0.60717\\\hline
3&0.99548&0.93171&0.7496&0.48539&0.22516&-0.001036&-0.18658&-0.33362&-0.44859&-0.53851\\\hline
3.5&0.99663&0.94794&0.80605&0.57728&0.33048&0.10724&-0.082998&-0.23907&-0.36446&-0.46448\\\hline
4&0.99744&0.95947&0.84831&0.65829&0.42809&0.21059&0.019383&-0.14256&-0.27634&-0.38544\\\hline
4.5&0.99802&0.96788&0.87915&0.72761&0.519&0.30886&0.11927&-0.045612&-0.18545&-0.30213\\\hline
5&0.99845&0.9742&0.90151&0.7837&0.60372&0.4027&0.21602&0.050386&-0.093273&-0.21578\\\hline
\end{tabular}
\end{table}
\begin{table}\begin{tabular}{|c|c|c|c|c|c|c|c|c|c|c|}
\hline
4&0.5&1&1.5&2&2.5&3&3.5&4&4.5&5\\\hline
0.5&0.98046&0.74081&0.33296&-0.040662&-0.29022&-0.46481&-0.58802&-0.67541&-0.73842&-0.78513\\\hline
1&0.98716&0.81882&0.46342&0.1203&-0.15253&-0.35437&-0.49811&-0.60151&-0.67727&-0.73396\\\hline
1.5&0.99122&0.87123&0.57267&0.24913&-0.027455&-0.24315&-0.40397&-0.52276&-0.61128&-0.67828\\\hline
2&0.9938&0.90659&0.66299&0.36351&0.09161&-0.13239&-0.30688&-0.43969&-0.54063&-0.61805\\\hline
2.5&0.9955&0.93079&0.73599&0.4646&0.20231&-0.024133&-0.2083&-0.35312&-0.4657&-0.55341\\\hline
3&0.99665&0.94766&0.79338&0.5537&0.3042&0.079939&-0.10978&-0.26405&-0.38704&-0.48457\\\hline
3.5&0.99746&0.95967&0.83742&0.63178&0.39761&0.17875&-0.012757&-0.17362&-0.3053&-0.41184\\\hline
4&0.99803&0.96837&0.87067&0.69917&0.48329&0.27188&0.081575&-0.082998&-0.22134&-0.3357\\\hline
4.5&0.99846&0.97481&0.89559&0.75576&0.56193&0.35953&0.17246&0.0067006&-0.13613&-0.25682\\\hline
5&0.99877&0.97969&0.91427&0.80156&0.63374&0.44227&0.25964&0.094581&-0.050699&-0.17605\\\hline
\end{tabular}
\end{table}
\begin{table}\begin{tabular}{|c|c|c|c|c|c|c|c|c|c|c|}
\hline
4.5&0.5&1&1.5&2&2.5&3&3.5&4&4.5&5\\\hline
0.5&0.98718&0.81502&0.44491&0.10032&-0.16945&-0.36159&-0.50701&-0.60895&-0.68285&-0.73835\\\hline
1&0.99123&0.86909&0.56173&0.23876&-0.038588&-0.24877&-0.41422&-0.53128&-0.61836&-0.68421\\\hline
1.5&0.99381&0.90543&0.6531&0.34983&0.077021&-0.14382&-0.31917&-0.45019&-0.54953&-0.6256\\\hline
2&0.9955&0.9302&0.72682&0.45018&0.18623&-0.040184&-0.22317&-0.36626&-0.4771&-0.56323\\\hline
2.5&0.99666&0.94744&0.78513&0.53819&0.28643&0.061467&-0.12762&-0.28041&-0.40159&-0.49735\\\hline
3&0.99746&0.9597&0.83032&0.615&0.37778&0.15769&-0.033839&-0.19365&-0.32365&-0.42832\\\hline
3.5&0.99804&0.96858&0.86489&0.68133&0.46089&0.24797&0.057062&-0.10704&-0.24403&-0.35657\\\hline
4&&0.97513&0.89123&0.73758&0.53647&0.33229&0.14431&-0.021558&-0.16354&-0.28266\\\hline
4.5&&0.98004&0.91134&0.7842&0.605&0.41103&0.22752&0.061979&-0.082998&-0.20721\\\hline
5&&0.9838&0.92676&0.82192&0.66655&0.48468&0.30669&0.143&-0.0031901&-0.13094\\\hline
\end{tabular}
\end{table}
\begin{table}\begin{tabular}{|c|c|c|c|c|c|c|c|c|c|c|}
\hline
5&0.5&1&1.5&2&2.5&3&3.5&4&4.5&5\\\hline
0.5&0.99124&0.86734&0.55698&0.20183&-0.059451&-0.27179&-0.42331&-0.5403&-0.62512&-0.68964\\\hline
1&0.99381&0.90456&0.64501&0.32805&0.058236&-0.15897&-0.32922&-0.45872&-0.55671&-0.63167\\\hline
1.5&0.99551&0.92979&0.71987&0.43811&0.1793&-0.052614&-0.2344&-0.37624&-0.48567&-0.57058\\\hline
2&0.99666&0.94729&0.77923&0.52678&0.27315&0.047679&-0.14077&-0.2923&-0.41206&-0.50647\\\hline
2.5&0.99747&0.95969&0.82543&0.60322&0.36351&0.14247&-0.048944&-0.20781&-0.33646&-0.43969\\\hline
3&&0.96866&0.86097&0.66921&0.44538&0.23113&0.039846&-0.12372&-0.25955&-0.37067\\\hline
3.5&&0.97528&0.88822&0.72538&0.51942&0.3136&0.12485&-0.040942&-0.18209&-0.29991\\\hline
4&&0.98023&0.90919&0.77236&0.58622&0.39015&0.20564&0.03975&-0.10481&-0.22798\\\hline
4.5&&0.984&0.92541&0.81098&0.64611&0.46118&0.2821&0.11779&-0.028406&-0.15547\\\hline
5&&0.9869&0.93807&0.84228&0.69917&0.52712&0.35437&0.19285&0.046535&-0.082998\\\hline
\end{tabular}
\end{table}
\clearpage
\subsection{The ratio between the inner and outer electron - $\tan\alpha$}
\clearpage
\begin{table}\begin{tabular}{|c|c|c|c|c|c|c|c|c|c|c|}
\hline
0.5&0.5&1&1.5&2&2.5&3&3.5&4&4.5&5\\\hline
0.5&1.1308&1.1356&1.1364&1.1367&1.1368&1.1369&1.1369&1.137&1.137&1.137\\\hline
1&1.4276&1.498&1.5451&1.5668&1.5777&1.584&1.5878&1.5903&1.5921&1.5933\\\hline
1.5&1.6387&2.1492&2.2456&2.2888&2.312&2.3257&2.3342&2.34&2.344&2.3469\\\hline
2&1.6425&2.186&3.2773&3.2658&3.2246&3.1996&3.1842&3.1743&3.1676&3.1628\\\hline
2.5&1.6264&2.1042&2.7489&4.7326&4.5053&4.3344&4.2261&4.1549&4.106&4.071\\\hline
3&1.6057&2.071&2.4545&3.4008&6.4124&5.9399&5.6162&5.4016&5.2541&5.1488\\\hline
3.5&1.5829&2.0519&2.3616&2.8151&3.9887&8.2731&7.5608&7.0716&6.7343&6.4942\\\hline
4&1.5584&2.0395&2.3142&2.6404&3.1691&10.1026&10.1111&9.3671&8.7044&8.2319\\\hline
4.5&1.5311&2.0309&2.2853&2.5532&2.9142&3.4869&4.6429&7.2826&10.1111&10.1111\\\hline
5&1.4977&2.0246&2.266&2.5007&2.7846&3.172&3.7527&4.7846&7.7297&8.1248\\\hline
\end{tabular}
\end{table}
\begin{table}\begin{tabular}{|c|c|c|c|c|c|c|c|c|c|c|}
\hline
1&0.5&1&1.5&2&2.5&3&3.5&4&4.5&5\\\hline
0.5&1.057&1.0599&1.0594&1.0591&1.059&1.0589&1.0588&1.0588&1.0588&1.0587\\\hline
1&1.1071&1.1308&1.1344&1.1356&1.1361&1.1364&1.1366&1.1367&1.1368&1.1368\\\hline
1.5&1.1719&1.243&1.2635&1.2772&1.2852&1.29&1.2931&1.2952&1.2966&1.2977\\\hline
2&1.2069&1.4276&1.46&1.498&1.5262&1.5451&1.5579&1.5668&1.5731&1.5777\\\hline
2.5&1.2048&1.5982&1.7347&1.7865&1.8322&1.8657&1.8894&1.9062&1.9184&1.9275\\\hline
3&1.1837&1.6387&1.9431&2.1492&2.2093&2.2456&2.2708&2.2888&2.3021&2.312\\\hline
3.5&1.1531&1.6455&1.9348&2.3209&2.6657&2.7138&2.726&2.7313&2.7346&2.7371\\\hline
4&1.1196&1.6425&1.9218&2.186&2.8158&3.2773&3.2894&3.2658&3.2429&3.2246\\\hline
4.5&1.0888&1.6354&1.9112&2.1334&2.4514&3.4861&3.9714&3.9273&3.8601&3.8031\\\hline
5&1.0647&1.6264&1.9024&2.1042&2.3363&2.7489&4.4053&4.7326&4.6211&4.5053\\\hline
\end{tabular}
\end{table}
\begin{table}\begin{tabular}{|c|c|c|c|c|c|c|c|c|c|c|}
\hline
1.5&0.5&1&1.5&2&2.5&3&3.5&4&4.5&5\\\hline
0.5&1.0341&1.0416&1.0413&1.041&1.0409&1.0408&1.0407&1.0407&1.0406&1.0406\\\hline
1&1.048&1.0785&1.0803&1.0803&1.0801&1.0799&1.0798&1.0798&1.0797&1.0797\\\hline
1.5&1.0603&1.1226&1.1308&1.1336&1.1349&1.1356&1.136&1.1363&1.1364&1.1365\\\hline
2&1.0717&1.1826&1.1994&1.2094&1.2162&1.2207&1.2237&1.2257&1.2271&1.2282\\\hline
2.5&1.0729&1.2621&1.2948&1.3124&1.3294&1.3428&1.3526&1.3598&1.3651&1.369\\\hline
3&1.0654&1.3252&1.4276&1.4472&1.4733&1.498&1.5179&1.5334&1.5451&1.5542\\\hline
3.5&1.0545&1.3558&1.5617&1.6205&1.6497&1.6822&1.7111&1.7346&1.7531&1.7676\\\hline
4&1.0437&1.3663&1.6192&1.7946&1.8611&1.9&1.9338&1.962&1.9846&2.0027\\\hline
4.5&1.0345&1.3656&1.6387&1.8413&2.0478&2.1492&2.1932&2.2231&2.2456&2.2634\\\hline
5&&1.3581&1.6449&1.8458&2.0453&2.3438&2.4826&2.5255&2.5461&2.559\\\hline
\end{tabular}
\end{table}
\begin{table}\begin{tabular}{|c|c|c|c|c|c|c|c|c|c|c|}
\hline
2&0.5&1&1.5&2&2.5&3&3.5&4&4.5&5\\\hline
0.5&1.0197&1.0321&1.0325&1.0323&1.0322&1.0321&1.032&1.0319&1.0319&1.0319\\\hline
1&1.0231&1.057&1.0599&1.0599&1.0596&1.0594&1.0592&1.0591&1.059&1.059\\\hline
1.5&1.0248&1.081&1.09&1.0916&1.0919&1.0919&1.0919&1.0918&1.0918&1.0917\\\hline
2&1.0272&1.1071&1.1258&1.1308&1.1331&1.1344&1.1351&1.1356&1.1359&1.1361\\\hline
2.5&1.0285&1.1389&1.1709&1.1801&1.1864&1.1908&1.1939&1.196&1.1976&1.1988\\\hline
3&1.0271&1.1719&1.2302&1.243&1.2538&1.2635&1.2712&1.2772&1.2817&1.2852\\\hline
3.5&1.024&1.1954&1.3038&1.3243&1.3378&1.353&1.367&1.3788&1.3883&1.3959\\\hline
4&1.0205&1.2069&1.3693&1.4276&1.4414&1.46&1.4798&1.498&1.5134&1.5262\\\hline
4.5&1.0172&1.2091&1.4098&1.5349&1.5679&1.5869&1.6101&1.6333&1.654&1.6717\\\hline
5&1.0143&1.2048&1.4312&1.5982&1.7022&1.7347&1.7604&1.7865&1.8108&1.8322\\\hline
\end{tabular}
\end{table}
\begin{table}\begin{tabular}{|c|c|c|c|c|c|c|c|c|c|c|}
\hline
2.5&0.5&1&1.5&2&2.5&3&3.5&4&4.5&5\\\hline
0.5&1.0106&1.0259&1.0268&1.0269&1.0268&1.0267&1.0266&1.0266&1.0266&1.0265\\\hline
1&1.0115&1.0439&1.0485&1.0488&1.0486&1.0484&1.0482&1.0481&1.048&1.0479\\\hline
1.5&1.0115&1.058&1.0698&1.0716&1.0718&1.0717&1.0715&1.0714&1.0713&1.0712\\\hline
2&1.0118&1.0704&1.0927&1.0973&1.0987&1.0992&1.0994&1.0995&1.0995&1.0995\\\hline
2.5&1.0123&1.0838&1.1186&1.1272&1.1308&1.1328&1.134&1.1347&1.1353&1.1356\\\hline
3&1.0123&1.099&1.1497&1.163&1.1693&1.1737&1.177&1.1793&1.1811&1.1824\\\hline
3.5&1.0117&1.1123&1.1874&1.2073&1.216&1.2233&1.2297&1.2348&1.2389&1.2421\\\hline
4&1.0106&1.1204&1.2273&1.2628&1.273&1.2828&1.2926&1.3016&1.3092&1.3155\\\hline
4.5&1.0094&1.1228&1.2601&1.3283&1.3432&1.3537&1.3664&1.3793&1.3911&1.4013\\\hline
5&1.0082&1.1206&1.2824&1.39&1.4276&1.4383&1.4521&1.468&1.4836&1.498\\\hline
\end{tabular}
\end{table}
\begin{table}\begin{tabular}{|c|c|c|c|c|c|c|c|c|c|c|}
\hline
3&0.5&1&1.5&2&2.5&3&3.5&4&4.5&5\\\hline
0.5&1.0057&1.0212&1.0227&1.023&1.023&1.0229&1.0229&1.0229&1.0228&1.0228\\\hline
1&1.0061&1.0341&1.0408&1.0416&1.0415&1.0413&1.0412&1.041&1.0409&1.0409\\\hline
1.5&1.0059&1.0422&1.057&1.0596&1.06&1.0599&1.0597&1.0595&1.0594&1.0593\\\hline
2&1.0059&1.048&1.073&1.0785&1.08&1.0803&1.0803&1.0803&1.0802&1.0801\\\hline
2.5&1.006&1.0537&1.0893&1.0993&1.1024&1.1037&1.1043&1.1046&1.1048&1.1048\\\hline
3&1.0061&1.0603&1.1071&1.1226&1.128&1.1308&1.1325&1.1336&1.1344&1.1349\\\hline
3.5&1.006&1.0668&1.1276&1.1499&1.1577&1.1624&1.1658&1.1683&1.1702&1.1717\\\hline
4&1.0058&1.0717&1.1503&1.1826&1.193&1.1994&1.2048&1.2094&1.2131&1.2162\\\hline
4.5&1.0054&1.0737&1.1719&1.2212&1.2354&1.243&1.2503&1.2572&1.2635&1.2688\\\hline
5&1.0049&1.0729&1.189&1.2621&1.2864&1.2948&1.3032&1.3124&1.3213&1.3294\\\hline
\end{tabular}
\end{table}
\begin{table}\begin{tabular}{|c|c|c|c|c|c|c|c|c|c|c|}
\hline
3.5&0.5&1&1.5&2&2.5&3&3.5&4&4.5&5\\\hline
0.5&1.0032&1.0175&1.0196&1.0199&1.02&1.0201&1.0201&1.02&1.02&1.02\\\hline
1&1.0034&1.0262&1.035&1.0363&1.0364&1.0363&1.0362&1.0361&1.036&1.0359\\\hline
1.5&1.0033&1.0306&1.0478&1.0512&1.0519&1.0519&1.0518&1.0516&1.0514&1.0513\\\hline
2&1.0032&1.0332&1.0592&1.0661&1.0679&1.0683&1.0683&1.0682&1.0681&1.068\\\hline
2.5&1.0032&1.0355&1.0698&1.0816&1.085&1.0862&1.0867&1.0868&1.0868&1.0868\\\hline
3&1.0033&1.0381&1.0804&1.098&1.1037&1.1061&1.1073&1.108&1.1083&1.1086\\\hline
3.5&1.0033&1.041&1.092&1.1161&1.1246&1.1285&1.1308&1.1323&1.1334&1.1341\\\hline
4&1.0033&1.0436&1.1049&1.1368&1.1485&1.1539&1.1576&1.1603&1.1624&1.164\\\hline
4.5&1.0032&1.045&1.1181&1.1605&1.1762&1.1832&1.1882&1.1924&1.1959&1.1989\\\hline
5&1.003&1.045&1.1297&1.1865&1.2088&1.2173&1.2234&1.229&1.2342&1.2389\\\hline
\end{tabular}
\end{table}
\begin{table}\begin{tabular}{|c|c|c|c|c|c|c|c|c|c|c|}
\hline
4&0.5&1&1.5&2&2.5&3&3.5&4&4.5&5\\\hline
0.5&1.0019&1.014&1.0171&1.0175&1.0177&1.0178&1.0178&1.0178&1.0178&1.0178\\\hline
1&1.002&1.0197&1.0303&1.0321&1.0325&1.0325&1.0324&1.0323&1.0322&1.0322\\\hline
1.5&1.002&1.022&1.0404&1.0449&1.0459&1.0461&1.046&1.0459&1.0457&1.0456\\\hline
2&1.0019&1.0231&1.0485&1.057&1.0593&1.0599&1.0599&1.0599&1.0597&1.0596\\\hline
2.5&1.0019&1.0239&1.0555&1.069&1.073&1.0744&1.0749&1.075&1.0749&1.0748\\\hline
3&1.0019&1.0248&1.0618&1.081&1.0876&1.09&1.0911&1.0916&1.0918&1.0919\\\hline
3.5&1.0019&1.026&1.0684&1.0936&1.1032&1.107&1.109&1.1101&1.1108&1.1112\\\hline
4&1.0019&1.0272&1.0757&1.1071&1.1203&1.1258&1.1288&1.1308&1.1322&1.1331\\\hline
4.5&1.0019&1.0282&1.0835&1.1222&1.1395&1.1469&1.1511&1.154&1.1563&1.1581\\\hline
5&1.0019&1.0285&1.0909&1.1389&1.1615&1.1709&1.1761&1.1801&1.1835&1.1864\\\hline
\end{tabular}
\end{table}
\begin{table}\begin{tabular}{|c|c|c|c|c|c|c|c|c|c|c|}
\hline
4.5&0.5&1&1.5&2&2.5&3&3.5&4&4.5&5\\\hline
0.5&1.0012&1.011&1.0151&1.0154&1.0157&1.0158&1.0159&1.016&1.016&1.016\\\hline
1&1.0013&1.0145&1.0263&1.0287&1.0293&1.0294&1.0294&1.0293&1.0293&1.0292\\\hline
1.5&1.0013&1.0158&1.0341&1.0398&1.0412&1.0416&1.0416&1.0415&1.0413&1.0412\\\hline
2&1.0012&1.0162&1.0399&1.0499&1.0527&1.0535&1.0537&1.0536&1.0535&1.0534\\\hline
2.5&1.0012&1.0164&1.0443&1.0592&1.0641&1.0658&1.0663&1.0664&1.0664&1.0663\\\hline
3&1.0012&1.0167&1.048&1.0682&1.0758&1.0785&1.0797&1.0801&1.0803&1.0803\\\hline
3.5&1.0012&1.0171&1.0518&1.077&1.0879&1.0921&1.094&1.095&1.0955&1.0958\\\hline
4&&1.0176&1.0558&1.0862&1.1007&1.1067&1.1096&1.1113&1.1123&1.1129\\\hline
4.5&&1.0181&1.0603&1.096&1.1145&1.1226&1.1266&1.1291&1.1308&1.132\\\hline
5&&1.0184&1.0648&1.1067&1.1299&1.1403&1.1455&1.1488&1.1513&1.1533\\\hline
\end{tabular}
\end{table}
\begin{table}\begin{tabular}{|c|c|c|c|c|c|c|c|c|c|c|}
\hline
5&0.5&1&1.5&2&2.5&3&3.5&4&4.5&5\\\hline
0.5&1.0007&1.0083&1.0134&1.0138&1.014&1.0142&1.0143&1.0144&1.0144&1.0145\\\hline
1&1.0008&1.0106&1.0227&1.0259&1.0266&1.0268&1.0269&1.0269&1.0268&1.0268\\\hline
1.5&1.0008&1.0113&1.0287&1.0356&1.0373&1.0379&1.038&1.0379&1.0378&1.0377\\\hline
2&1.0008&1.0115&1.0326&1.0439&1.0474&1.0485&1.0488&1.0488&1.0487&1.0486\\\hline
2.5&1.0008&1.0115&1.0354&1.0513&1.057&1.059&1.0597&1.0599&1.0599&1.0599\\\hline
3&&1.0115&1.0375&1.058&1.0666&1.0698&1.0711&1.0716&1.0718&1.0718\\\hline
3.5&&1.0116&1.0396&1.0642&1.0762&1.081&1.083&1.084&1.0845&1.0847\\\hline
4&&1.0118&1.0418&1.0704&1.0859&1.0927&1.0957&1.0973&1.0982&1.0987\\\hline
4.5&&1.0121&1.0442&1.0768&1.0962&1.1051&1.1093&1.1117&1.1131&1.114\\\hline
5&&1.0123&1.0468&1.0838&1.1071&1.1186&1.1241&1.1272&1.1293&1.1308\\\hline
\end{tabular}
\end{table}
\clearpage
\subsection{Energy in a.u.}
\clearpage
\begin{table}\begin{tabular}{|c|c|c|c|c|c|c|c|c|c|c|}
\hline
0.5&0.5&1&1.5&2&2.5&3&3.5&4&4.5&5\\\hline
0.5&-15.8981&-16.0481&-16.0395&-16.0347&-16.0321&-16.0307&-16.0298&-16.0292&-16.0288&-16.0285\\\hline
1&-9.6104&-9.7523&-9.8078&-9.8254&-9.8333&-9.8375&-9.84&-9.8416&-9.8428&-9.8435\\\hline
1.5&-12.1324&-9.1606&-8.9144&-8.9215&-8.9335&-8.9418&-8.9474&-8.9513&-8.954&-8.956\\\hline
2&-11.6194&-15.3448&-8.9021&-8.7354&-8.7145&-8.7117&-8.7127&-8.7144&-8.716&-8.7173\\\hline
2.5&-10.1134&-19.9432&-12.3688&-8.6712&-8.6078&-8.5968&-8.595&-8.5957&-8.5969&-8.5983\\\hline
3&-8.5262&-22.2207&-16.7013&-10.7812&-8.5156&-8.5024&-8.504&-8.5075&-8.511&-8.514\\\hline
3.5&-7.0664&-23.4749&-19.4534&-13.9316&-10.1196&-8.4113&-8.4184&-8.4277&-8.4356&-8.442\\\hline
4&-5.7598&-24.212&-21.3162&-16.2227&-12.3743&-8.4276&-8.2991&-8.3521&-8.3651&-8.3757\\\hline
4.5&-4.5811&-24.6675&-22.6298&-17.9354&-14.1399&-11.5127&-9.7526&-9.0442&-8.0489&-8.2237\\\hline
5&-3.4563&-24.9611&-23.587&-19.2394&-15.5553&-12.8701&-11.0121&-9.7165&-8.8509&-9.2044\\\hline
\end{tabular}
\end{table}
\begin{table}\begin{tabular}{|c|c|c|c|c|c|c|c|c|c|c|}
\hline
1&0.5&1&1.5&2&2.5&3&3.5&4&4.5&5\\\hline
0.5&-6.208&-6.6877&-6.7206&-6.7285&-6.7316&-6.7332&-6.7341&-6.7347&-6.7351&-6.7353\\\hline
1&-3.271&-3.9745&-4.0109&-4.012&-4.0109&-4.0099&-4.0092&-4.0087&-4.0083&-4.008\\\hline
1.5&-1.9851&-2.809&-2.8827&-2.8896&-2.8892&-2.8882&-2.8874&-2.8868&-2.8863&-2.8859\\\hline
2&-1.3088&-2.4026&-2.4172&-2.4381&-2.4473&-2.4519&-2.4546&-2.4564&-2.4575&-2.4583\\\hline
2.5&-0.80624&-2.7744&-2.3259&-2.2788&-2.2831&-2.2892&-2.2937&-2.2969&-2.2992&-2.3009\\\hline
3&-0.40925&-3.0331&-3.0319&-2.2901&-2.2355&-2.2286&-2.2288&-2.2304&-2.232&-2.2334\\\hline
3.5&-0.098585&-3.0303&-3.9577&-2.7741&-2.2591&-2.2134&-2.2026&-2.1996&-2.1989&-2.199\\\hline
4&0.14091&-2.9049&-4.4501&-3.8362&-2.5247&-2.2255&-2.1935&-2.1838&-2.1801&-2.1786\\\hline
4.5&0.32165&-2.7263&-4.7228&-4.5213&-3.4383&-2.3587&-2.1942&-2.1742&-2.1672&-2.1642\\\hline
5&0.45509&-2.5283&-4.8724&-4.9858&-4.1248&-3.0922&-2.2451&-2.1678&-2.1562&-2.1519\\\hline
\end{tabular}
\end{table}
\begin{table}\begin{tabular}{|c|c|c|c|c|c|c|c|c|c|c|}
\hline
1.5&0.5&1&1.5&2&2.5&3&3.5&4&4.5&5\\\hline
0.5&-3.022&-3.6505&-3.7138&-3.7273&-3.7326&-3.7352&-3.7367&-3.7377&-3.7383&-3.7388\\\hline
1&-1.6078&-2.3693&-2.4439&-2.4555&-2.4586&-2.4598&-2.4603&-2.4606&-2.4608&-2.4609\\\hline
1.5&-0.83231&-1.6621&-1.7665&-1.7811&-1.7831&-1.7831&-1.7828&-1.7824&-1.7822&-1.7819\\\hline
2&-0.37643&-1.2496&-1.377&-1.4008&-1.4046&-1.4048&-1.4042&-1.4037&-1.4032&-1.4028\\\hline
2.5&-0.077294&-1.0364&-1.154&-1.184&-1.1928&-1.1955&-1.1964&-1.1967&-1.1968&-1.1968\\\hline
3&0.13849&-0.93321&-1.0678&-1.0693&-1.0783&-1.0836&-1.0866&-1.0885&-1.0898&-1.0906\\\hline
3.5&0.30084&-0.84323&-1.1599&-1.0425&-1.0255&-1.027&-1.03&-1.0324&-1.0343&-1.0357\\\hline
4&0.42493&-0.74597&-1.2894&-1.191&-1.0273&-1.0063&-1.0035&-1.004&-1.0052&-1.0064\\\hline
4.5&0.52061&-0.64578&-1.348&-1.494&-1.1476&-1.0178&-0.99703&-0.99178&-0.99049&-0.99044\\\hline
5&&-0.54797&-1.3555&-1.6926&-1.499&-1.1002&-1.0088&-0.99048&-0.98448&-0.9822\\\hline
\end{tabular}
\end{table}
\begin{table}\begin{tabular}{|c|c|c|c|c|c|c|c|c|c|c|}
\hline
2&0.5&1&1.5&2&2.5&3&3.5&4&4.5&5\\\hline
0.5&-1.5649&-2.2644&-2.3443&-2.3624&-2.3692&-2.3725&-2.3744&-2.3756&-2.3764&-2.377\\\hline
1&-0.78746&-1.552&-1.6525&-1.6719&-1.6777&-1.6802&-1.6814&-1.6821&-1.6826&-1.6829\\\hline
1.5&-0.31622&-1.1091&-1.2333&-1.2576&-1.2635&-1.2653&-1.2659&-1.2662&-1.2664&-1.2665\\\hline
2&-0.01252&-0.81776&-0.96151&-0.99363&-1.001&-1.0027&-1.0031&-1.003&-1.0029&-1.0027\\\hline
2.5&0.19438&-0.62421&-0.77911&-0.81891&-0.82948&-0.83209&-0.83255&-0.83243&-0.83217&-0.83191\\\hline
3&0.34342&-0.49627&-0.66047&-0.70226&-0.71617&-0.72067&-0.72205&-0.7224&-0.72241&-0.72231\\\hline
3.5&0.45547&-0.40346&-0.59796&-0.62938&-0.64268&-0.64875&-0.65148&-0.65278&-0.65345&-0.65383\\\hline
4&0.54201&-0.32721&-0.57628&-0.60065&-0.60006&-0.60429&-0.60746&-0.60952&-0.61088&-0.61182\\\hline
4.5&0.61005&-0.26066&-0.56261&-0.63155&-0.58956&-0.58149&-0.58195&-0.58351&-0.58497&-0.58616\\\hline
5&0.66433&-0.20156&-0.54206&-0.69359&-0.63629&-0.58147&-0.5713&-0.56969&-0.57001&-0.57078\\\hline
\end{tabular}
\end{table}
\begin{table}\begin{tabular}{|c|c|c|c|c|c|c|c|c|c|c|}
\hline
2.5&0.5&1&1.5&2&2.5&3&3.5&4&4.5&5\\\hline
0.5&-0.77889&-1.5172&-1.6046&-1.6263&-1.6345&-1.6384&-1.6406&-1.6419&-1.6428&-1.6434\\\hline
1&-0.30845&-1.0675&-1.1845&-1.2106&-1.2187&-1.2221&-1.2238&-1.2248&-1.2255&-1.2259\\\hline
1.5&-0.0024938&-0.77359&-0.9079&-0.9397&-0.94863&-0.95181&-0.95321&-0.95394&-0.95436&-0.95463\\\hline
2&0.20694&-0.56781&-0.71551&-0.75414&-0.76503&-0.76845&-0.76968&-0.77018&-0.7704&-0.7705\\\hline
2.5&0.35632&-0.42078&-0.57711&-0.62219&-0.63592&-0.64016&-0.64149&-0.64188&-0.64196&-0.64192\\\hline
3&0.4667&-0.31477&-0.47668&-0.52626&-0.54304&-0.54867&-0.55049&-0.551&-0.55107&-0.55098\\\hline
3.5&0.55091&-0.23622&-0.406&-0.45686&-0.47563&-0.48285&-0.48554&-0.48648&-0.48676&-0.48678\\\hline
4&0.61682&-0.17473&-0.35846&-0.40995&-0.42765&-0.43572&-0.43934&-0.44095&-0.44167&-0.44199\\\hline
4.5&0.6694&-0.1241&-0.32436&-0.38529&-0.39665&-0.40318&-0.40704&-0.40919&-0.41041&-0.41112\\\hline
5&0.71198&-0.081109&-0.29559&-0.37974&-0.38442&-0.38358&-0.3857&-0.38766&-0.38909&-0.39009\\\hline
\end{tabular}
\end{table}
\begin{table}\begin{tabular}{|c|c|c|c|c|c|c|c|c|c|c|}
\hline
3&0.5&1&1.5&2&2.5&3&3.5&4&4.5&5\\\hline
0.5&-0.30643&-1.0438&-1.1509&-1.1941&-1.1933&-1.1978&-1.2002&-1.2017&-1.2027&-1.2034\\\hline
1&-0.00069879&-0.75549&-0.88236&-0.91262&-0.92422&-0.92844&-0.93058&-0.93182&-0.93261&-0.93314\\\hline
1.5&0.20909&-0.55092&-0.68978&-0.72713&-0.73873&-0.74308&-0.74506&-0.74611&-0.74674&-0.74714\\\hline
2&0.35911&-0.40195&-0.54931&-0.59233&-0.6061&-0.61096&-0.61295&-0.61388&-0.61437&-0.61465\\\hline
2.5&0.47001&-0.29137&-0.44386&-0.49195&-0.50833&-0.51413&-0.51632&-0.51723&-0.51762&-0.51781\\\hline
3&0.55426&-0.20808&-0.36345&-0.41553&-0.43454&-0.44161&-0.44427&-0.44528&-0.44566&-0.44578\\\hline
3.5&0.6198&-0.14428&-0.30218&-0.35684&-0.378&-0.38648&-0.38984&-0.39114&-0.39161&-0.39175\\\hline
4&0.67188&-0.094108&-0.25594&-0.3124&-0.33461&-0.34425&-0.34843&-0.35019&-0.3509&-0.35116\\\hline
4.5&0.71397&-0.053357&-0.22057&-0.28041&-0.30211&-0.31211&-0.31694&-0.31923&-0.3203&-0.32078\\\hline
5&0.74853&-0.019323&-0.19199&-0.25911&-0.27979&-0.2885&-0.29335&-0.296&-0.29743&-0.2982\\\hline
\end{tabular}
\end{table}
\begin{table}\begin{tabular}{|c|c|c|c|c|c|c|c|c|c|c|}
\hline
3.5&0.5&1&1.5&2&2.5&3&3.5&4&4.5&5\\\hline
0.5&-0.00010259&-0.75117&-0.86452&-0.89078&-0.89958&-0.90596&-0.9153&-0.91693&-0.91799&-0.91873\\\hline
1&0.20963&-0.54274&-0.67529&-0.71221&-0.72347&-0.72852&-0.73104&-0.73248&-0.73339&-0.734\\\hline
1.5&0.35971&-0.39494&-0.53534&-0.57663&-0.59054&-0.59597&-0.59847&-0.59981&-0.6006&-0.60111\\\hline
2&0.47075&-0.28422&-0.42982&-0.47562&-0.4917&-0.49783&-0.50047&-0.50176&-0.50247&-0.5029\\\hline
2.5&0.55519&-0.19979&-0.34816&-0.39785&-0.41624&-0.42337&-0.42632&-0.42766&-0.42833&-0.4287\\\hline
3&0.62085&-0.13435&-0.28403&-0.33678&-0.35738&-0.36571&-0.36917&-0.37066&-0.37135&-0.37167\\\hline
3.5&0.67296&-0.082841&-0.23335&-0.28827&-0.31076&-0.32034&-0.32445&-0.32623&-0.327&-0.32734\\\hline
4&0.71496&-0.041623&-0.19327&-0.2497&-0.27356&-0.28425&-0.28909&-0.29126&-0.29221&-0.29262\\\hline
4.5&0.74938&-0.0079718&-0.16141&-0.21943&-0.24401&-0.25545&-0.26095&-0.26358&-0.2648&-0.26536\\\hline
5&0.77792&0.020072&-0.13551&-0.19618&-0.22114&-0.23269&-0.23861&-0.24165&-0.24319&-0.24395\\\hline
\end{tabular}
\end{table}
\begin{table}\begin{tabular}{|c|c|c|c|c|c|c|c|c|c|c|}
\hline
4&0.5&1&1.5&2&2.5&3&3.5&4&4.5&5\\\hline
0.5&0.20982&-0.5376&-0.66212&-0.70654&-0.71465&-0.71777&-0.72024&-0.72211&-0.72324&-0.72404\\\hline
1&0.35991&-0.39122&-0.52604&-0.5661&-0.57987&-0.58606&-0.58897&-0.59061&-0.59163&-0.59231\\\hline
1.5&0.47095&-0.28119&-0.42177&-0.46569&-0.48158&-0.48801&-0.491&-0.49259&-0.49354&-0.49414\\\hline
2&0.55541&-0.19687&-0.34054&-0.388&-0.40588&-0.41312&-0.41636&-0.41798&-0.41888&-0.41943\\\hline
2.5&0.62113&-0.13113&-0.27622&-0.32661&-0.34646&-0.35472&-0.35835&-0.36008&-0.36098&-0.36149\\\hline
3&0.67329&-0.079056&-0.22468&-0.27729&-0.29897&-0.30832&-0.31247&-0.3144&-0.31536&-0.31586\\\hline
3.5&0.71538&-0.037211&-0.18307&-0.2372&-0.26045&-0.2709&-0.27568&-0.2779&-0.27897&-0.2795\\\hline
4&0.74966&-0.0031301&-0.1493&-0.20444&-0.22894&-0.24038&-0.24581&-0.24841&-0.24965&-0.25025\\\hline
4.5&0.77828&0.025013&-0.12176&-0.17771&-0.20311&-0.21532&-0.22138&-0.22439&-0.22587&-0.22659\\\hline
5&0.80221&0.048594&-0.099048&-0.15605&-0.18208&-0.19478&-0.20132&-0.20473&-0.20648&-0.20737\\\hline
\end{tabular}
\end{table}
\begin{table}\begin{tabular}{|c|c|c|c|c|c|c|c|c|c|c|}
\hline
4.5&0.5&1&1.5&2&2.5&3&3.5&4&4.5&5\\\hline
0.5&0.35996&-0.38839&-0.52139&-0.55926&-0.57246&-0.57615&-0.58095&-0.583&-0.58386&-0.5845\\\hline
1&0.471&-0.2795&-0.41622&-0.45738&-0.47342&-0.47881&-0.48414&-0.48598&-0.48711&-0.48786\\\hline
1.5&0.55551&-0.19551&-0.33577&-0.3814&-0.39885&-0.40561&-0.40971&-0.41156&-0.41264&-0.41334\\\hline
2&0.62125&-0.12985&-0.27192&-0.32031&-0.33955&-0.34776&-0.35156&-0.35348&-0.35456&-0.35522\\\hline
2.5&0.67338&-0.077704&-0.22049&-0.27103&-0.29192&-0.3011&-0.30533&-0.30742&-0.30853&-0.30918\\\hline
3&0.71549&-0.035657&-0.17864&-0.23072&-0.2531&-0.26326&-0.26802&-0.27034&-0.27154&-0.2722\\\hline
3.5&0.7499&-0.0013094&-0.14433&-0.1974&-0.22105&-0.23215&-0.23749&-0.24012&-0.24145&-0.24216\\\hline
4&&0.027084&-0.116&-0.16967&-0.19435&-0.20631&-0.21224&-0.21522&-0.21674&-0.21753\\\hline
4.5&&0.050815&-0.092479&-0.14655&-0.17201&-0.18468&-0.19117&-0.19453&-0.19627&-0.19718\\\hline
5&&0.070869&-0.072789&-0.12729&-0.15332&-0.16653&-0.1735&-0.17723&-0.17922&-0.18027\\\hline
\end{tabular}
\end{table}
\begin{table}\begin{tabular}{|c|c|c|c|c|c|c|c|c|c|c|}
\hline
5&0.5&1&1.5&2&2.5&3&3.5&4&4.5&5\\\hline
0.5&0.47105&-0.27812&-0.40882&-0.45856&-0.46914&-0.47611&-0.47791&-0.48063&-0.48157&-0.48232\\\hline
1&0.55551&-0.19472&-0.33222&-0.37931&-0.395&-0.40114&-0.40454&-0.40657&-0.40781&-0.40863\\\hline
1.5&0.62125&-0.12922&-0.26908&-0.31604&-0.33321&-0.34269&-0.34661&-0.34874&-0.34997&-0.35075\\\hline
2&0.67346&-0.077113&-0.21802&-0.26687&-0.28709&-0.29613&-0.30044&-0.30266&-0.30391&-0.30468\\\hline
2.5&0.71549&-0.035045&-0.17632&-0.22672&-0.24832&-0.25824&-0.263&-0.26542&-0.26673&-0.26751\\\hline
3&&-0.00062345&-0.14197&-0.1934&-0.2162&-0.22697&-0.23225&-0.23492&-0.23635&-0.23716\\\hline
3.5&&0.027882&-0.11345&-0.16548&-0.18929&-0.20088&-0.20669&-0.20967&-0.21124&-0.21211\\\hline
4&&0.051736&-0.089595&-0.14195&-0.16657&-0.17888&-0.18522&-0.18854&-0.1903&-0.19126\\\hline
4.5&&0.071894&-0.069513&-0.12205&-0.14727&-0.1602&-0.16704&-0.1707&-0.17268&-0.17376\\\hline
5&&0.089079&-0.052488&-0.1052&-0.13084&-0.14428&-0.15155&-0.15555&-0.15776&-0.15898\\\hline
\end{tabular}
\end{table}

\clearpage
\subsection{Residuals}
\clearpage
\begin{table}\begin{tabular}{|c|c|c|c|c|c|c|c|c|c|c|}
\hline
0.5&0.5&1&1.5&2&2.5&3&3.5&4&4.5&5\\\hline
0.5&1.7038e-07&3.0378e-07&3.4241e-07&5.3881e-07&6.675e-07&1.0543e-06&5.1811e-07&7.1297e-07&2.6575e-06&1.7004e-06\\\hline
1&7.5684e-08&1.0492e-08&1.3892e-08&9.8446e-08&3.7037e-07&3.195e-07&5.599e-07&6.0841e-07&6.5192e-07&1.1575e-06\\\hline
1.5&2.856e-07&3.9541e-08&6.533e-08&1.5955e-07&3.1663e-08&2.5868e-07&3.4383e-07&3.3726e-07&4.727e-07&5.652e-07\\\hline
2&5.7869e-07&1.1331e-07&4.4958e-08&6.1134e-08&1.0663e-07&1.9839e-07&2.0661e-07&3.3351e-07&2.5003e-08&4.9501e-07\\\hline
2.5&7.3297e-07&2.1436e-07&6.4963e-08&8.8881e-09&6.0161e-08&4.7534e-08&1.4161e-07&2.1489e-07&3.1082e-08&2.0563e-07\\\hline
3&3.2364e-06&4.1367e-08&1.6944e-07&5.484e-08&6.0086e-09&2.9451e-08&1.1469e-07&1.5416e-07&1.8259e-07&2.7833e-07\\\hline
3.5&2.0554e-06&5.9847e-07&2.6514e-07&1.0133e-07&2.5778e-08&4.9825e-08&1.025e-08&3.4986e-08&5.9008e-09&1.1312e-07\\\hline
4&1.2122e-05&6.338e-07&3.7363e-07&2.5415e-07&1.262e-07&0.00060675&0.0079052&5.8231e-08&8.2798e-08&1.0762e-07\\\hline
4.5&2.682e-06&6.185e-07&5.643e-07&3.7106e-07&2.2394e-07&1.6003e-07&4.3583e-05&0.041401&0.052446&0.018366\\\hline
5&0.0015594&1.2656e-06&3.2661e-07&2.2325e-07&3.5845e-07&1.7876e-07&0.00019721&4.3024e-06&0.018797&0.080306\\\hline
\end{tabular}
\end{table}
\begin{table}\begin{tabular}{|c|c|c|c|c|c|c|c|c|c|c|}
\hline
1&0.5&1&1.5&2&2.5&3&3.5&4&4.5&5\\\hline
0.5&1.3872e-06&2.772e-06&1.9435e-06&1.2862e-06&3.4569e-06&1.4843e-06&2.5998e-06&3.3037e-06&3.2027e-06&3.5289e-06\\\hline
1&3.8657e-07&6.8152e-07&6.7352e-07&1.2151e-06&1.2532e-06&1.3696e-06&8.4549e-07&2.1552e-06&1.25e-06&2.67e-06\\\hline
1.5&1.8044e-07&7.5652e-07&5.6404e-07&3.3405e-07&6.9498e-07&4.5473e-07&3.8647e-07&1.72e-06&1.7088e-06&2.2171e-06\\\hline
2&1.0486e-06&3.0273e-07&4.7602e-07&4.1969e-08&1.2785e-07&5.5568e-08&6.2876e-07&3.9379e-07&8.9376e-07&1.4815e-06\\\hline
2.5&2.1373e-06&2.3489e-07&1.7174e-07&2.3497e-07&3.24e-07&2.8586e-07&4.2231e-07&3.6458e-07&2.8237e-07&5.4156e-07\\\hline
3&3.5255e-06&1.1424e-06&2.0921e-07&1.5816e-07&7.5377e-08&2.6132e-07&4.9865e-07&6.3821e-07&6.6683e-07&1.2665e-07\\\hline
3.5&6.5278e-06&1.8497e-06&3.5494e-07&9.3057e-08&1.0389e-07&1.7286e-07&2.6782e-07&1.6796e-07&4.3437e-07&3.1263e-07\\\hline
4&7.9631e-06&2.3147e-06&1.0094e-06&4.5325e-07&1.1376e-07&1.7983e-07&2.3908e-07&2.4454e-07&2.8529e-07&4.265e-07\\\hline
4.5&2.4819e-05&3.0575e-06&8.1075e-07&6.2852e-07&1.8543e-07&7.9253e-08&8.3181e-08&1.4564e-07&4.0721e-07&1.0702e-07\\\hline
5&5.804e-05&2.9319e-06&1.4802e-06&8.5743e-07&6.4994e-08&2.5985e-07&4.5353e-08&3.5552e-08&2.7278e-07&2.4065e-07\\\hline
\end{tabular}
\end{table}
\begin{table}\begin{tabular}{|c|c|c|c|c|c|c|c|c|c|c|}
\hline
1.5&0.5&1&1.5&2&2.5&3&3.5&4&4.5&5\\\hline
0.5&5.9883e-05&0.0075714&3.2797e-06&4.258e-06&1.5687e-06&6.6698e-06&1.8129e-06&6.6491e-06&6.8731e-06&3.4914e-06\\\hline
1&2.9309e-06&2.8199e-06&5.7779e-06&6.3096e-07&3.7589e-06&4.5457e-06&4.6941e-06&1.6395e-06&2.2999e-06&5.8928e-06\\\hline
1.5&4.2713e-06&2.8035e-06&1.5334e-06&3.6309e-06&2.3562e-06&2.734e-06&3.186e-06&2.6274e-06&3.0817e-06&2.4111e-06\\\hline
2&8.252e-06&2.2538e-06&9.3892e-07&1.1433e-06&9.1586e-07&2.1223e-06&1.3972e-06&1.4677e-06&1.5475e-06&2.7692e-06\\\hline
2.5&5.8316e-06&7.3107e-07&1.6066e-06&8.0548e-07&1.013e-06&1.1806e-06&1.273e-06&1.2035e-06&1.2563e-06&2.3337e-06\\\hline
3&1.7773e-05&2.2298e-06&6.8115e-07&8.0632e-07&5.2685e-07&9.443e-08&3.6139e-07&1.0531e-06&1.2503e-07&9.4652e-07\\\hline
3.5&3.3652e-05&3.7812e-06&1.0277e-06&6.2875e-07&8.6913e-07&6.1085e-07&6.2872e-07&4.111e-07&9.0016e-07&6.7788e-07\\\hline
4&5.3942e-05&5.9354e-06&1.2485e-06&4.7627e-07&3.3403e-07&5.6266e-07&7.8219e-07&6.8347e-07&3.7986e-07&1.3351e-06\\\hline
4.5&8.5812e-05&7.4505e-06&2.5704e-06&6.4943e-07&2.2794e-07&3.5586e-07&4.4502e-07&4.9123e-07&5.8797e-07&7.2925e-08\\\hline
5&&2.8523e-06&1.7521e-06&1.0786e-06&5.8137e-07&1.4234e-07&2.7442e-07&2.9903e-07&1.7228e-07&6.0144e-07\\\hline
\end{tabular}
\end{table}
\begin{table}\begin{tabular}{|c|c|c|c|c|c|c|c|c|c|c|}
\hline
2&0.5&1&1.5&2&2.5&3&3.5&4&4.5&5\\\hline
0.5&7.0712e-06&2.9192e-05&9.5084e-06&8.9413e-06&3.1656e-06&4.7336e-06&1.4165e-05&2.2096e-06&1.7196e-05&1.3097e-05\\\hline
1&8.6509e-06&5.5489e-06&1.286e-05&1.1088e-05&3.4246e-06&7.7741e-06&1.3058e-06&5.1447e-06&4.1627e-06&1.3828e-05\\\hline
1.5&1.6749e-05&1.893e-06&1.2059e-06&7.037e-06&7.0362e-06&8.0383e-06&4.9907e-06&8.9169e-06&2.5477e-06&7.3605e-06\\\hline
2&1.5022e-05&1.5463e-06&6.287e-06&2.7261e-06&5.0407e-06&2.6941e-06&2.2283e-06&4.8604e-06&4.6126e-06&5.0129e-06\\\hline
2.5&2.2672e-05&3.8767e-06&2.1918e-06&8.8741e-07&2.7385e-06&1.5293e-06&1.1236e-06&3.7669e-06&3.553e-06&3.9441e-06\\\hline
3&5.3741e-05&7.2176e-07&3.4909e-06&3.0261e-06&1.0626e-06&2.2562e-06&1.6493e-06&1.3362e-06&1.6478e-06&2.7799e-06\\\hline
3.5&9.4679e-05&4.331e-06&2.2955e-06&1.5444e-06&1.322e-06&9.5763e-07&2.1287e-06&1.1368e-06&8.1383e-07&9.8677e-07\\\hline
4&0.00010715&4.1945e-06&2.2838e-06&1.2109e-06&1.7675e-06&1.9041e-06&1.3354e-06&1.6788e-07&1.2108e-06&5.1141e-07\\\hline
4.5&0.00023066&3.8672e-06&2.1638e-06&6.4361e-07&2.7997e-07&1.3527e-06&9.3374e-07&2.4682e-07&1.5894e-06&9.7251e-07\\\hline
5&8.687e-05&8.5492e-06&4.7259e-06&9.3955e-07&3.1931e-07&6.8695e-07&7.7173e-07&9.3986e-07&1.2468e-06&1.296e-06\\\hline
\end{tabular}
\end{table}
\begin{table}\begin{tabular}{|c|c|c|c|c|c|c|c|c|c|c|}
\hline
2.5&0.5&1&1.5&2&2.5&3&3.5&4&4.5&5\\\hline
0.5&1.4241e-05&0.070643&0.0077937&2.6946e-05&4.0637e-06&3.1766e-05&1.1023e-05&7.2968e-06&3.7419e-05&6.4739e-06\\\hline
1&1.1169e-05&1.7796e-05&9.3869e-06&2.1949e-05&2.2877e-05&2.4567e-05&2.1389e-05&9.348e-06&1.05e-05&8.777e-06\\\hline
1.5&2.2974e-05&1.1401e-05&1.4549e-05&1.4139e-05&3.1871e-06&1.3748e-05&6.2859e-06&2.0646e-06&2.3547e-06&5.3816e-06\\\hline
2&1.5444e-05&8.3381e-06&1.141e-05&4.3233e-06&1.6249e-06&4.3699e-06&4.3728e-06&9.3171e-06&1.4094e-06&8.9004e-06\\\hline
2.5&1.0914e-05&5.4918e-06&7.6961e-06&9.8267e-06&4.2595e-06&7.1158e-07&8.169e-06&8.1679e-06&8.0915e-06&7.5944e-06\\\hline
3&7.3021e-05&7.9849e-06&8.5056e-06&3.5026e-06&6.0784e-06&7.7112e-06&2.4307e-06&5.622e-06&4.4814e-06&4.3178e-06\\\hline
3.5&0.00017081&1.255e-05&4.9868e-06&5.7947e-06&2.4232e-06&1.715e-06&8.0812e-07&4.0467e-06&4.9869e-06&3.4789e-06\\\hline
4&0.00022304&9.9074e-06&4.45e-06&4.4702e-06&3.843e-06&3.9772e-06&3.5456e-06&1.9843e-06&2.3532e-06&3.197e-06\\\hline
4.5&0.00037635&2.2597e-05&6.0539e-06&2.5577e-06&2.6539e-06&1.2036e-06&9.9348e-07&3.4159e-06&2.4699e-06&2.122e-06\\\hline
5&0.00057159&3.2038e-05&1.0071e-05&3.5606e-06&1.8921e-06&2.4446e-06&2.8565e-06&8.1188e-07&1.3486e-06&2.6231e-07\\\hline
\end{tabular}
\end{table}
\begin{table}\begin{tabular}{|c|c|c|c|c|c|c|c|c|c|c|}
\hline
3&0.5&1&1.5&2&2.5&3&3.5&4&4.5&5\\\hline
0.5&3.446e-05&0.0045926&0.063704&0.13834&3.3034e-05&7.0137e-06&3.7289e-05&7.7121e-07&6.8398e-06&4.6164e-05\\\hline
1&9.7495e-06&0.00023953&0.0029661&0.030286&3.4502e-05&1.3119e-05&4.5022e-06&1.7032e-05&7.8121e-06&6.275e-06\\\hline
1.5&7.7742e-05&1.9186e-05&1.2485e-05&2.1805e-05&1.1956e-05&2.4948e-05&2.8818e-05&2.0552e-05&1.7492e-05&1.4733e-05\\\hline
2&7.1978e-05&1.1724e-05&9.4702e-06&1.128e-05&1.4101e-05&2.3111e-05&1.611e-05&2.5238e-06&5.7132e-06&1.5036e-05\\\hline
2.5&0.00015196&7.0302e-06&1.6968e-05&1.7866e-05&1.8081e-05&1.2221e-05&1.2569e-05&1.8457e-05&1.1464e-05&1.0182e-05\\\hline
3&0.00013023&1.7085e-05&3.4791e-06&1.1214e-05&1.182e-05&6.1337e-06&2.6155e-06&1.4524e-05&6.0617e-06&9.4249e-06\\\hline
3.5&0.00025751&8.5425e-06&4.8597e-06&6.6204e-06&6.9798e-06&3.6422e-06&1.0687e-05&7.6372e-06&8.7789e-06&1.0597e-05\\\hline
4&0.00037258&3.3008e-05&7.4118e-06&9.0153e-06&1.388e-06&3.7557e-06&2.2542e-06&4.5731e-06&4.4887e-06&3.6635e-06\\\hline
4.5&0.00068795&1.5671e-05&1.624e-06&6.8064e-06&8.0708e-06&6.8087e-06&3.4695e-06&2.5133e-06&5.0764e-06&3.5483e-06\\\hline
5&0.00048974&2.3326e-05&8.8884e-06&2.9243e-06&5.6596e-06&6.4266e-06&3.9151e-06&3.2219e-06&4.4898e-06&4.0519e-06\\\hline
\end{tabular}
\end{table}
\begin{table}\begin{tabular}{|c|c|c|c|c|c|c|c|c|c|c|}
\hline
3.5&0.5&1&1.5&2&2.5&3&3.5&4&4.5&5\\\hline
0.5&0.00013317&0.11243&0.0031309&0.12389&0.1865&0.15824&6.4003e-05&5.0429e-05&6.132e-05&5.8011e-05\\\hline
1&0.00015729&4.3182e-05&0.013738&0.031233&2.2928e-05&3.9389e-05&3.8946e-05&1.0904e-05&5.8163e-05&2.8758e-05\\\hline
1.5&0.00011046&8.4819e-06&2.9048e-05&4.0013e-05&0.00020079&2.8134e-05&4.5278e-05&2.2186e-05&1.695e-05&4.7722e-06\\\hline
2&0.00014184&4.6402e-06&3.0793e-05&2.3715e-05&1.5616e-05&2.3099e-05&3.5691e-05&1.7603e-05&1.433e-05&1.2884e-05\\\hline
2.5&0.00018125&2.8299e-05&8.2184e-06&2.9666e-06&2.8079e-05&2.0076e-05&1.8107e-05&1.1128e-05&4.5354e-06&8.1867e-06\\\hline
3&6.2129e-05&3.9461e-05&1.0143e-05&1.8059e-05&1.1313e-05&9.9825e-06&1.6257e-05&2.3141e-05&2.076e-05&2.0541e-05\\\hline
3.5&0.0006388&2.0728e-05&1.0257e-05&1.7666e-05&5.543e-06&1.113e-05&8.3486e-06&6.2246e-06&1.8767e-05&6.5879e-06\\\hline
4&0.00065054&3.036e-05&1.3843e-05&5.7358e-06&1.583e-05&4.7555e-06&1.4559e-05&8.8762e-06&6.3386e-06&1.9768e-06\\\hline
4.5&0.0010228&2.4041e-05&1.3377e-05&7.3917e-06&8.5521e-06&1.3535e-05&3.1206e-06&1.8998e-06&1.0498e-05&3.0531e-06\\\hline
5&0.00078802&4.0884e-05&1.7881e-05&6.6755e-06&6.2515e-06&2.6892e-06&5.7638e-06&1.0814e-05&6.4587e-06&7.9712e-06\\\hline
\end{tabular}%
\end{table}
\end{document}